\documentclass[amsmath,showpacs,aps,prb,twocolumn,floatfix]{revtex4-1}
\usepackage{bm}
\usepackage{graphics}
\usepackage{graphicx}
\usepackage{amsthm}
\usepackage{amsmath}
\usepackage{wasysym}
\usepackage{amssymb}
\usepackage{enumerate}
\usepackage{xfrac}
\usepackage[dvips]{epsfig}
\usepackage[caption=false]{subfig}
\usepackage[normalem]{ulem}
\usepackage{color}
\usepackage{txfonts}
\usepackage[mathscr]{euscript}
\usepackage{hyperref} 
\usepackage{verbatim}

\begin{document}
\title{Thermoelectric properties of Weyl and Dirac semimetals}
\author{Rex Lundgren}
\email{rexlund@physics.utexas.edu}
\author{Pontus Laurell}
\author{Gregory A. Fiete}
\affiliation{Department of Physics, The University of Texas at Austin, Austin, TX 78712, USA}
\date{\today}
\pacs{71.10.-w,71.20.-b,72.15.Jf}

 

\begin{abstract}
We study the electronic contribution to the thermal conductivity and the thermopower of Weyl and Dirac semimetals using a semiclassical Boltzmann approach.  We investigate the effect of various relaxation processes including disorder and interactions on the thermoelectric properties, and also consider doping away from the Weyl or Dirac point.  We find that the thermal conductivity and thermopower have an interesting dependence on the chemical potential that is characteristic of the linear electronic dispersion, and that the electron-electron interactions modify the Lorenz number. For the interacting system, we also use the Kubo formalism to obtain the transport coefficients. We find exact agreement between the Kubo and Boltzmann approaches  at high temperatures. We also consider the effect of electric and magnetic fields on the thermal conductivity in various orientations with respect to the temperature gradient. Notably, when the temperature gradient and magnetic field are parallel, we find a large contribution to the longitudinal thermal conductivity that is quadratic in the magnetic field strength, similar to the magnetic field dependence of the longitudinal electrical conductivity due to the presence of the chiral anomaly when no thermal gradient is present.
\end{abstract}
\maketitle

\section{Introduction}
Dirac and Weyl semimetals have enjoyed a surge of interest since the prediction of their existence at the phase transition between a three-dimensional topological insulator \cite{PhysRevLett.96.106802} and a normal insulator,\cite{2007NJPh....9..356M}  and in pyrochlore iridates,\cite{PhysRevB.83.205101} respectively. Dirac (Weyl) semimetals have linearly dispersing excitations [which obey the Dirac (Weyl) equation] from degenerate band touching points referred to as Dirac (Weyl) nodes.  The electronic states around the band degeneracy points possess a non-zero Berry curvature,\cite{Volovik} which gives rise to non-trivial momentum space topology.  Weyl nodes are separated in momentum space, always come in pairs of opposite chirality,\cite{1983PhLB..130..389N} and act like magnetic monopoles in momentum space with quantized Berry flux.  Because the two touching bands must be expressed locally in momentum space in terms of the complete basis of Pauli matrices, there is no perturbation that can be added to open a gap; the momentum space location of the Weyl point only shifts slightly, leading to a ``topologically protected" metallic phase. Dirac semimetals can be thought of as two Weyl nodes of different chirality not separated in momentum space. Dirac points can be protected by symmetry, but are generally unstable compared to the Weyl semimetal.\cite{PhysRevLett.108.140405}

Other topological aspects of Weyl semimetals include the chiral (or Adler-Bell-Jackiw) anomaly,\cite{PhysRev.177.2426,1969NCimA..60...47B} chiral magnetic effect,\cite{PhysRevD.78.074033,2014PrPNP..75..133K} and edge states referred to as Fermi arcs.\cite{PhysRevB.83.205101,PhysRevB.86.195102,PhysRevB.89.235315,2014arXiv1401.0529H,2014arXiv1402.6342P,PhysRevB.84.245415,2014arXiv1406.3804L} We now briefly describe these topological effects. When parallel electric and magnetic fields are applied, the axial current (difference between currents of different chirality)  is not conserved. The chiral anomaly is the mathematical statement that the number of particles with a given chirality is not conserved. Recent attempts to find a theoretical description of an experimental smoking gun signature of the chiral anomaly include predictions for optical phenomena, \cite{PhysRevB.87.245131,PhysRevB.87.235121,2014arXiv1401.2762H,PhysRevB.89.245121,2014arXiv1404.2927G} density response, \cite{PhysRevB.89.245103} electromagnetic response, \cite{PhysRevLett.111.027201} and non-local transport. \cite{PhysRevX.4.031035}  If a pair of Weyl nodes have band touching points at different energies, a chiral magnetic effect -- the separation of electric charge along the direction of an external magnetic field -- occurs.\cite{PhysRevD.78.074033,2014PrPNP..75..133K}  One interesting application of the chiral magnetic effect is chiral electronics, which refers to circuits with elements that take advantage of the chiral current that arises due to the external magnetic field.\cite{PhysRevB.88.115119}  Such systems have been proposed as quantum amplifiers of magnetic fields.  Finally, Weyl semimetals possess Fermi arcs on certain physical edges of a crystal\cite{2014arXiv1401.0529H} that can give rise to novel structure in Friedel oscillations in thin-film systems.\cite{PhysRevB.86.195102,2014arXiv1402.6342P} For an overview of the physics of Weyl semimetals from a condensed matter perspective, see the reviews by Turner and Vishwanath,\cite{2013arXiv1301.0330T} and Hosur and Qi.\cite{2013CRPhy..14..857H} 

There is compelling experimental evidence for the Dirac semimetal phase in $\mathrm{Na}_3\mathrm{Bi}$,\cite{Liu21022014,2013arXiv1312.7624X} and $\mathrm{Cd}_3\mathrm{As}_2$.\cite{2014NatCo...5E3786N,PhysRevLett.113.027603,2014arXiv1403.3446J,2014arXiv1404.2557H} Numerical calculations predict the phase also to occur in $A_3 \mathrm{Bi}$, where $A=\mathrm{K, Rb}$,\cite{PhysRevB.85.195320} and in $\mathrm{BiO}_2$.\cite{PhysRevLett.108.140405}  Starting from a Dirac semimetal, it is possible to obtain a Weyl semimetal phase by breaking time reversal or inversion symmetry.\cite{Xu29042011,PhysRevLett.109.186403,2014arXiv1403.3446J,PhysRevB.88.165105,PhysRevB.85.035103,PhysRevB.85.165110}  The minimum number of Weyl nodes for time reversal (inversion) symmetry broken realizations of Weyl semimetals is 2 (4). The Weyl semimetal phase has also been predicted to appear in a wide range of materials in addition to the previously mentioned pyrochlore iridates.\cite{PhysRevB.83.205101,PhysRevB.84.075129,PhysRevB.87.155101,PhysRevB.85.045124,annurev-conmatphys-020911-125138} These include: stacked layers of three dimensional topological insulators and normal insulators,\cite{PhysRevLett.107.127205} ferromagnetic compounds such as HgC$\mathrm{r}_2$S$\mathrm{e}_4$,\cite{PhysRevLett.107.186806} $\mathrm{Hg}_{1-\mathrm{x}-\mathrm{y}}\mathrm{Cd}_\mathrm{x}\mathrm{Mn}_\mathrm{y}\mathrm{Te}$ wells,\cite{PhysRevB.89.081106} gyroid photonic crystals,\cite{2013NaPho...7..294L} orbital-selective superlattices,\cite{PhysRevB.88.035444} optical lattices,\cite{2014arXiv1405.4866G} and Fulde-Ferrell superfluids.\cite{PhysRevLett.112.136402} Interesting phases that can be induced from Weyl semimetals by interactions include excitonic phases,\cite{PhysRevB.89.235109} superconductivity,\cite{PhysRevB.86.214514,2014arXiv1405.4299H} spin density waves,\cite{PhysRevB.90.035126} charge density waves and axion strings.\cite{PhysRevB.87.161107,2014arXiv1406.4501R} Other phases closely related to Weyl semimetals include the line-node semimetal,\cite{PhysRevB.84.235126} Weyl superconductors,\cite{PhysRevB.86.054504,2013arXiv1312.3632G,PhysRevLett.113.046401} Weyl semimetals with $\mathbb{Z}_2$ topological charge,\cite{PhysRevB.89.235127} and fractionalized Weyl semimetals.\cite{2014arXiv1405.0842W,2014arXiv1406.0843W}

There has been much work on the electrical transport properties of Dirac and Weyl semimetals,\cite{1983PhLB..130..389N,PhysRevB.33.3257,PhysRevB.33.3263,PhysRevLett.108.046602,PhysRevB.84.235126,PhysRevB.85.241101,2012JHEP...04..097L,PhysRevB.89.054202,PhysRevB.88.045108,PhysRevB.87.205133,PhysRevB.87.155123,PhysRevB.89.085126,PhysRevB.87.155123,1367-2630-15-12-123019,PhysRevB.89.155104,PhysRevLett.113.026602,PhysRevB.89.075124,2014arXiv1402.3737S,PhysRevB.89.245110,PhysRevLett.112.016402,PhysRevB.89.245110} with a particular focus on the case when the Fermi energy lies at the Dirac or Weyl node. One of the possible fingerprints of the chiral anomaly, first pointed out by Nielsen and Ninomiya,\cite{1983PhLB..130..389N} is negative longitudinal magnetoresistance (decreasing longitudinal resistance with increasing magnetic field). Negative magnetoresistance has already been observed in a Dirac semimetal with broken time reversal symmetry, suggesting a possible Weyl semimetal phase.\cite{PhysRevLett.111.246603} Another interesting transport feature of Weyl semimetals is that the anomalous Hall conductivity depends linearly on the distance in momentum space between Weyl nodes,\cite{PhysRevLett.107.127205} whereas it vanishes for Dirac semimetals.

In this paper we investigate the electronic contribution to the thermal conductivity and thermopower of Dirac and Weyl semimetals. We use the Boltzmann equation to analytically calculate the thermoelectric coefficients for various relaxation processes including short-range disorder, scattering off charged impurities, which also change the Fermi energy of the system, and electron-electron interactions. For electrical transport in Weyl semimetals, the Boltzmann approach equation is in good agreement with other theoretical approaches such as the Kubo formula and the quantum Boltzmann equation. For example, (ignoring rare region effects \cite{PhysRevB.89.245110}) the Boltzmann equation and the Kubo formula give the same result for the electrical conductivity in disordered systems in the dc limit. \cite{PhysRevLett.108.046602,PhysRevB.84.235126} Furthermore, the Boltzmann equation gives exactly the same rate of change of the number of particles of a given chirality as relativistic quantum field theories. \cite{PhysRevB.88.104412} A Boltzmann equation approach has also proved useful for understanding transport in graphene,\cite{PhysRevB.76.073412}  which may serve as a two-dimensional analog, in some respects, to a Weyl or Dirac semimetal. 

For the relaxation processes we considered, there is an interesting dependence of the thermal conductivity and thermopower on the Fermi level due to the relativistic dispersion relation. These results are summarized in Table~\ref{Nofields}. When the relaxation processes are dominated by interactions, we also find an interesting quadratic temperature dependence for the thermal conductivity compared to a linear temperature dependence when scattering is dominated by charged impurities or disorder. We also investigate the effect of electric and magnetic fields on the thermoelectric coefficients. In the absence of an electric field, we find that for a magnetic field perpendicular to the temperature gradient the transverse thermal conductivity is linear in magnetic field strength (for small fields) and the longitudinal thermal conductivity has a quadratic field dependence that {\em decreases } the magnitude of the thermal conductivity. When the magnetic field is parallel to the temperature gradient we find a quadratic magnetic field dependence for the longitudinal thermal conductivity that {\em increases} its magnitude and zero transverse thermal conductivity. When electric fields are present, we find no additional transport terms  compared to all situations in which it is zero (essentially resulting from the assumption of linear response), due to the cancellation of terms with different chirality. The results for the thermal conductivity in the presence of electric and magnetic fields is summarized in Table~\ref{Fields} and should be helpful in identifying three-dimensional systems with Dirac and Weyl points. We also calculate thermoelectric coefficients via the Kubo formula and compare our results obtained from the Boltzmann equation and find exact agreement between the two approaches at high temperatures.

Our paper is organized as follows. In Sec.~\ref{sec:TTC}, we solve the Boltzmann equation to obtain transport coefficients for various relaxation processes including electron-electron interactions, charged impurities which also change the Fermi energy, and short-range disorder. In Sec.~\ref{sec:TTCEB}, we investigate the effect of electric and magnetic fields on thermoelectric transport properties. In Sec.~\ref{sec:Kubo}, for the case of interacting electrons, we use the Kubo formula to calculate thermoelectric transport coefficients and compare to results obtained from the semiclassical approach. Finally, in Sec.~\ref{sec:Con}, we present the main conclusions of our paper.  A few technical results are relegated to the appendices.
\section{Thermoelectric Transport Coefficients} \label{sec:TTC}
\subsection{Formalism and Anomalous Transport}
In this section we investigate the electronic contributions to the thermal conductivity and thermopower in the absence of a magnetic field of a single Weyl node, with a given chirality, described by the following Hamiltonian
\begin{equation}
H=\chi\hbar v_f \vec{\sigma}\cdot \vec{k},
\end{equation}
where $\chi$ is the chirality that takes values of $\pm$ 1, $v_f$ is the Fermi velocity, $\vec{\sigma}=\{\sigma_x,\sigma_y,\sigma_z\}$ is a vector of Pauli matrices, and $\vec{k}$ is the wavevector. Our fundamental physical conclusions are not changed if the velocity is different in different directions, though asymmetric transport properties will result.  To generalize to an arbitrary number of Weyl nodes, one simply adds the conductivity for each node together, provided one is able to ignore inter-node scattering which may open a gap.\cite{PhysRevB.89.245110}  If the Fourier component of the scattering potential is ``small" at the wave vector connecting the nodes, this can be safely done.  While disorder may gap a Dirac node (two Weyl nodes ``on top" of each other), if the Fermi energy is larger than the gap induced in the node the system will remain metallic with an approximately linear dispersion.  We will assume this is the case throughout this work.  For relaxation through electron-electron interactions, zero wave vector scattering is insensitive to chirality\cite{PhysRevLett.108.046602} and as such, our results for electron-electron interactions apply to both Dirac and Weyl semimetals. To obtain the thermoelectric coefficients for a single Dirac node, one must add the results of two Weyl nodes with opposite chirality together. 

In our work, we ignore contributions to the thermoelectric coefficients from phonons, which are expected to dominate thermal transport above the Debye temperature.\cite{Ashcroft} In this situation, one would add the phonon (or even magnon) contributions to the electronic one to find the total thermal conductivity. For pyrochlore iridates, the first material predicted to host a Weyl semimetal phase, changing the rare-earth element will have little effect on the Debye temperature. Representative values are 420 K for $\mathrm{Eu}_2\mathrm{Ir}_2\mathrm{0}_7$,\cite{doi:10.1143/JPSJ.70.2880} and 400K for $\mathrm{Y}_2\mathrm{Ir}_2\mathrm{0}_7$.\cite{doi:10.1143/JPSJ.71.2578} These temperatures should be compared to the magnetic ordering temperature, which is on the order of 120 K for pyrochlore iridates. \cite{annurev-conmatphys-020911-125138} Hence, there is a large separation of the characteristic energy scales for magnetism and phonons, implying that the coupling between them is not too strong (otherwise, one would expect them to be more comparable in magnitude).  Moreover, in the pyrochlore iridates, the magnetism is crucial to the Weyl phase itself because one must break either inversion symmetry or time-reversal symmetry, and the former is preserved in the absence of magnetism.  Furthermore, the theoretical description of the Weyl phase assumes that magnetic fluctuations are not too strong, implying that our description of the thermal transport should apply at temperature sufficiently low that they can be ignored.  This condition restricts the validity of our treatment to be somewhat below the magnetic ordering temperature, which is itself significantly below the Debye temperature.  Thus, phonons should have little effect on thermal transport in the Weyl semimetal phase of the pyrochlore iridates. Generically, we expect electron-phonon interactions to give rise to a small correction to the electronic contributions to the thermal conductivity given the relatively large difference between the Debye temperature and the magnetic ordering temperature seen in most materials.\cite{annurev-conmatphys-020911-125138,doi:10.1143/JPSJ.70.2880,doi:10.1143/JPSJ.71.2578,Ashcroft}  On the other hand, electron-phonon interactions can have other interesting effects, such as leading to the appearance of a Weyl semimetal at non-zero temperatures from material that is an insulator at zero temperature.\cite{PhysRevLett.110.046402} For systems when the Debye temperature and magnetic ordering temperature are comparable, the arguments above may not hold.  We leave a detailed consideration of that situation to other work.

We now introduce the formalism for our semi-classical Boltzmann equation approach. In terms of the notation of Refs.~[\onlinecite{Ashcroft}] and ~[\onlinecite{PhysRevLett.97.026603}], the electrical current for a given chirality $\chi$ of a single Weyl node is
\begin{eqnarray}
\vec{J}^{\chi}=-e\int\frac{\mathrm{d}^3k}{(2\pi)^3}\left( \vec{v} + \frac{e}{\hbar}\vec{E}\times\vec{\Omega}^{\chi} \right) f^{\chi}+\hspace{3cm}\nonumber \\
\frac{\vec \nabla T}{T}\times\left(\frac{e}{\hbar}\int\frac{\mathrm{d}^3k}{(2\pi)^3}\vec{\Omega}^{\chi}\{(\epsilon-\mu)f_{eq}+k_B T\mathrm{log}(1+e^{-\beta(\epsilon-\mu^{\chi})})\}\right),\nonumber \\
\label{electric_current}
\end{eqnarray}
where $e$ is the electrical charge, $\hbar$ is Planck's constant divided by $2\pi$, $\mu$ is the chemical potential, $\epsilon=\hbar v_f k$ is the dispersion for quasiparticles with wavenumber $k$, $\vec{v}=\nabla_{\vec k} \epsilon$ is the semiclassical velocity, $\vec{E}$ is the electric field, $\beta=\frac{1}{k_BT}$, where $k_B$ is the Boltzmann constant and $T$ is the temperature, $f_{eq}$ is the Fermi-Dirac distribution, and $f^{\chi}$ is the quantum distribution function of the system that must be computed from the Boltzmann equation given below. $\vec \Omega^{\chi}=\vec \nabla_k\times \vec A^{\chi}_k$ is the Berry curvature and $\vec A_k^{\chi}=i\langle u^{\chi}_k|\vec \nabla_ k| u_k^{\chi} \rangle$ is the Berry connection with the Bloch eigenstate $|u_k^{\chi}\rangle$. The Berry curvature is proportional to $\chi$, so carries an opposite sign for Weyl nodes of opposite chirality and is therefore zero for a Dirac node. For simplicity, the chemical potential is assumed to be the same for all nodes.

The second term in Eq.~\eqref{electric_current} gives rise to the anomalous Nernst effect (non-zero transverse electric field in the absence of a magnetic field) and arises due to the finite spread of the wave packet.\cite{PhysRevLett.97.026603} From Refs.~[\onlinecite{Ashcroft}], [\onlinecite{PhysRevLett.97.026603}] and [\onlinecite{PhysRevLett.107.236601}], the thermal current is
\begin{align}
\vec{J}_q^{\chi}=\vec{J}_E^{\chi}-\mu\vec{J}^{\chi}=
\int\frac{\mathrm{d}^3k}{(2\pi)^3}(\epsilon-\mu)\frac{\partial \epsilon}{\partial \vec{k}} f^{\chi}\hspace{2.8cm}\nonumber \\
+\int\frac{\mathrm{d}^3k}{(2\pi)^3}\left(\vec{E}\times\frac{e}{\hbar}\vec{\Omega}^{\chi}\{(\epsilon-\mu)f_{eq}+k_BT\mathrm{log}(1+e^{-\beta(\epsilon-\mu)})\}\right)\nonumber \\
+\frac{\vec \nabla T}{T}\times\frac{e}{\hbar}\int\frac{\mathrm{d}^3k}{(2\pi)^3}\vec{\Omega}^{\chi}(\epsilon-\mu)^2f_{eq},
\label{Thermal_current}
\end{align}
where $\vec{J}_E^{\chi}=\int\frac{\mathrm{d}^3k}{(2\pi)^3}f\epsilon\vec{v}$ is the energy current. The first term is the standard expression for energy current in the absence of Berry curvature.\cite{Ashcroft} As before, the second term in Eq.~\eqref{Thermal_current} arises from the finite size spread of the wave-packet and gives rise to a transverse thermal current from the external electric field due to Berry curvature. \cite{PhysRevLett.97.026603} The last term in Eq.~\eqref{Thermal_current} gives rise to the anomalous thermal Hall effect (transverse thermal current from thermal gradient due to Berry curvature).\cite{PhysRevLett.107.236601} The results for the anomalous Nernst and anomalous thermal Hall effect apply also for interacting systems. \cite{PhysRevLett.93.206602} Alternatively, one could obtain these results by adding a pseudo-gravitational potential which acts like a temperature gradient. \cite{PhysRev.135.A1505,PhysRevLett.107.236601}

We obtain $f^{\chi}$ by solving the Boltzmann equation,\cite{Ashcroft,PhysRevLett.97.026603,PhysRevLett.107.236601} which is given by (in the absence of magnetic fields, which will be treated later)
\begin{align}
\frac{\partial{f}^{\chi}}{\partial t}+\{(\vec{v}+e\vec{E}\times\vec{\Omega}^{\chi})\cdot\vec \nabla_r f^{\chi}+
e\vec{E}\cdot\vec \nabla_k f^{\chi}\}=\mathrm{I}_{coll}^{\chi},
\end{align}
where $\mathrm{I}_{coll}^\chi$ is the collision integral at the Weyl node with chirality $\chi$. The temperature gradient (which we take to define the $x$-direction) and electric field are taken to be in the $x-$direction, which allows us to drop the $\vec{E}\times\vec{\Omega}^{\chi}$ term  since the spatial gradient of the distribution function is parallel to the thermal gradient. We solve the Boltzmann equation, via the relaxation time approximation, \cite{Ashcroft} in which case the collision integral takes the form $\mathrm{I}_{coll}^{\chi}=-\frac{f^{\chi}-f_{eq}}{\tau}$, where $\tau(k)$ is the intra-node scattering time. Following Ref.~[\onlinecite{Ashcroft}], we assume the following steady-state solution for the distribution function
\begin{equation}
f^{\chi}=f_{eq}+\tau(\epsilon(k))\left(-\frac{\partial f_{eq}}{\partial \epsilon}\right) \vec{v} \cdot \left( -e\vec{E}+\frac{\epsilon(k)-\mu}{T}(-\vec \nabla T)\right),
\end{equation}
valid in the linear response regime. From there, we can write
\begin{align}
J_{\alpha}^{\chi}=L^{11}_{\alpha\beta}E_{\beta}+L^{12}_{\alpha\beta}(-\vec \nabla_{\beta} T), \\
J_{q,\alpha}^{\chi}=L^{21}_{\alpha\beta}E_{\beta}+L^{22}_{\alpha\beta}(-\vec \nabla_{\beta} T),
\end{align}
where $\alpha$ and $\beta$ are spatial indices running over $x,y,z$, and the set of $L$ are the transport coefficients we are interested in obtaining. We will focus on longitudinal transport first, {\it i.e.}, $\alpha=\beta$. For transport along the electric field and thermal gradient (for simplicity, we assume a uniform temperature gradient and electric field), we have
$L^{11}_{xx}=\sigma_{xx}=\mathcal{L}^{0}$, $L^{21}_{xx}=TL^{12}_{xx}=-\mathcal{L}^{1}/e$ and $L^{22}_{xx}=\mathcal{L}^{2}/e^2T$, where 
\begin{equation}
\mathcal{L}^{\alpha}=e^2\int\mathrm{d}\epsilon\left(-\frac{\partial f^\chi}{\partial \epsilon}\right) \tau(\epsilon) g(\epsilon)v_x^2(\epsilon-\mu)^{\alpha},
\label{mathL}
\end{equation}
where $g(\epsilon)$ is the density of states, which (for Weyl and Dirac semimetals  when multiplied by a factor of 2) is given by
\begin{equation}
g(\epsilon)=\frac{\epsilon^2}{2\pi^2\hbar^3v_f^3},
\end{equation}
and $v_x$ is given by
\begin{equation}
v_x=\frac{\partial \epsilon}{\partial(\hbar k_x)}= v_f\frac{k_x}{|\vec{k}|}.
\end{equation}
With knowledge of the $\mathcal{L}^{\alpha}$, we can calculate the thermal conductivity (defined when no electrical current flows) as
\begin{equation}
\kappa_{xx}=L^{22}_{xx}-L_{xx}^{21}(L^{11}_{xx})^{-1}L^{12}_{xx},
\end{equation} 
and the Seebeck coefficient (or thermopower), which is given by
\begin{equation}
S=\frac{L^{12}_{xx}}{L^{11}_{xx}}.
\end{equation}
This approach is valid as long as the quasiparticle energy is much greater than the scattering rate. Longitudinal transport, in the absence of a magnetic field, is independent of the chirality since the $\chi$ dependence drops out of Eq.\eqref{mathL}.

Before moving on to specific forms of the scattering time, we briefly review results on transport transverse to the electric field and temperature gradient. The presence of Berry curvature introduces anomalous transport, {\it i.e.} transport in the transverse direction of the electric field and/or temperature gradient. For anomalous transport at low temperatures we have, from Ref.~[\onlinecite{PhysRevLett.97.026603}] and Ref.~[\onlinecite{PhysRevLett.107.236601}],
\begin{align}
\sigma_{\alpha x}=-\varepsilon_{\alpha x l}\frac{e^2}{\hbar}\int\frac{\mathrm{d}^3k}{(2\pi)^3}\Omega^{\chi}_l f_{eq},
\end{align}
\begin{align}
L^{12}_{\alpha x}=TL^{21}_{\alpha x}=-\varepsilon_{\alpha x l}\frac{1}{T}\frac{e}{\hbar}\int\frac{\mathrm{d}^3k}{(2\pi)^3}\Omega_l^{\chi}(\epsilon-\mu)f_{eq},
\end{align}
and
\begin{align}
L^{22}_{\alpha x}=-\varepsilon_{\alpha x l}\frac{1}{T}\frac{1}{\hbar}\int\frac{\mathrm{d}^3k}{(2\pi)^3}\Omega_l^{\chi}(\epsilon-\mu)^2f_{eq},
\end{align}
where $\alpha$ is either in the $y$ or $z$ direction and $\varepsilon$ is the Levi-Civita symbol.  Einstein summation is assumed. As we see, anomalous transport is determined only by the Berry curvature and band structure, as opposed to longitudinal transport which depends crucially on the details of the relaxation process. The anomalous transport coefficients obey the Mott relation $L^{12}_{\alpha x}=\frac{\pi^2}{3e}k_B T\frac{ \partial  \sigma_{\alpha x}}{\partial \mu}$, and the Wiedemann-Franz law, $\frac{\kappa_{\alpha x}}{T\sigma_{\alpha x}}=\frac{\pi^2}{3e^2}k_B^2 $, at low temperatures.\cite{PhysRevLett.97.026603,PhysRevLett.107.236601} For Weyl semimetals with only broken inversion symmetry, the anomalous Hall terms vanish. This is because when one groups Weyl nodes into pairs of opposite chirality, the resulting sum of the separation vectors between Weyl nodes vanishes. In contrast, for time-reversal symmetry broken realizations of Weyl semimetals, the anomalous Hall terms are non-zero.\cite{PhysRevB.87.245112} This offers a way to distinguish between time reversal and inversion symmetry broken realizations of Weyl semimetals. The anomalous electrical conductivity of time reversal symmetry broken realizations of Weyl semimetals when the Fermi energy is at the Weyl nodes has been studied in Ref.~[\onlinecite{PhysRevLett.107.127205}] and was found to be
\begin{equation}
 \sigma_{\alpha x}=-\varepsilon_{\alpha x l}\frac{e^2}{h}\frac{\Delta k_l}{\pi},
\label{anaelectricalcon}
\end{equation}
where $\Delta k_l$ is the distance in momentum space between Weyl nodes in the $l$th direction. This is because two-dimensional slices of Brillouin zone perpendicular to the direction between Weyl nodes are two-dimensional Chern insulators, which exhibit a quantized Hall effect.\cite{PhysRevLett.107.127205} In between these Weyl nodes, the two-dimensional slices of the Brillouin zone have non-trivial Chern number. Adding up the transverse conductivity of each slice gives Eq.~\eqref{anaelectricalcon}. When the Fermi energy is away from the Weyl nodes, the same results hold for an unbounded linear quasiparticle dispersion.\cite{PhysRevB.88.245107,2014arXiv1406.3033B} Furthermore (again only for unbounded linear dispersion), this result holds for finite temperatures. \cite{PhysRevD.9.3320,finite_Temp_anomaly} In real materials one expects small non-universal corrections to Eq.~\eqref{anaelectricalcon} due to band curvature effects. Using the Mott relation and the Wiedemann-Franz law, we obtain
\begin{equation}
L^{12}_{\alpha x}=-\frac{\pi^2}{3e}k_B T\frac{ \partial  \sigma_{\alpha x}}{\partial \mu}=0,
\label{Nernst_Weyl}
\end{equation}
and
\begin{equation}
L^{22}_{\alpha x}=-\frac{\pi}{3h}k_B^2T\varepsilon_{\alpha x l}\Delta k_l.
\label{Thermal_Hall_1}
\end{equation}
To our knowledge, Eq.~\eqref{Nernst_Weyl} describing the Anomalous Nernst Effect in Weyl semimetals, has not been previously obtained. The thermal Hall effect, described by Eq.~\eqref{Thermal_Hall_1}, was studied in Ref. [\onlinecite{PhysRevB.88.245107}] from a field theory point-of-view where it was found that the thermal Hall conductivity depends linearly on the distance between Weyl nodes, in agreement with Eq.~\eqref{Thermal_Hall_1}. Anomalous transport vanishes for Dirac semimetals due to the presence of time reversal symmetry \cite{PhysRevB.84.235126} and can vanish for Weyl semimetals in systems with cubic symmetry.\cite{PhysRevB.84.075129}

\subsection{Charged Impurities}

We first calculate the transport coefficients for scattering off charged impurities, which lead to dopants in the band structure and move the Fermi level away from the nodal point. The transport time was computed in the first Born approximation in the work of Burkov, Hook, and Balents (BHB).\cite{PhysRevB.84.235126}  They used a screened Coulomb potential given by
\begin{equation}
V(q)=\frac{4\pi e^2}{\epsilon_d (q^2+q_{TF}^2)},
\end{equation}
where $q_{TF}^2=\frac{4\pi e^2}{\epsilon_d} g(\epsilon)$ is the Thomas-Fermi wave vector and $\epsilon_d$ is the background dielectric constant. We note that a more accurate dielectric function with logarithmic corrections has been worked out in Ref.~[\onlinecite{DielectricWSM}] by evaluating the polarization bubble. However, in the following we will neglect those corrections. BHB found the scattering time to be
\begin{equation}
\frac{1}{\tau_{\rm screened}(\epsilon)}=\frac{4\pi^3 n_i \hbar^2v_f^3}{3\epsilon^2}f(\alpha),
\label{eq:tau_doping}
\end{equation}
where $n_i$ is the density of charged impurities and
\begin{equation}
f(\alpha)=\frac{3\alpha^2}{\pi^2}\left\{(1+\alpha/\pi)\mathrm{atanh}\left(\frac{1}{1+\alpha/\pi}\right)-1\right\},
\end{equation}
where $\alpha=\frac{e^2}{\epsilon_d v_f}$ and is the ratio of Coulomb (potential) energy to kinetic energy. We assume the charged impurities act as donors,  so that $\epsilon_f \propto v_f n_i^{1/3}$. $f(\alpha)$ arises from using Fermi's golden rule to calculate the transport time.  Physically it measures the strength of the interaction between electrons and the charge impurities. Our approach is valid when $f(\alpha)$ and $\alpha$ is small (which is simply the condition that the inverse transport time be less than the Fermi energy) and $n_i>0$. In the limit of small $\alpha$, $f(\alpha)\approx\frac{3\alpha^2}{2\pi^2}\mathrm{ln}(\alpha^{-1})$. Using Eq.~\eqref{eq:tau_doping} as the scattering time, and evaluating the integral in Eq.~\eqref{mathL} (e.g. using the Sommerfeld expansion), we find
\begin{align}
	\sigma_{xx}		&=	\frac{e^2 \epsilon_f}{8\pi^5 v_f \hbar^2 f(\alpha )} \left( 1 + 2\pi^2 \left( \frac{k_B T}{\epsilon_f}\right)^2 + \frac{7\pi^4}{15} \left(\frac{k_B T}{\epsilon_f}\right)^4 \right),\\
	L^{12}_{xx}		&=	-\frac{ek_B^2 T}{6\pi^3 \hbar^2 v_f f(\alpha)} \left( 1 + \frac{7\pi^2}{5} \left(\frac{ k_B T}{\epsilon_f}\right)^2  \right),\\
	L^{22}_{xx}		&=	\frac{k_B^2 T\epsilon_f}{24\pi^3\hbar^2v_f f(\alpha)} \left(1+\frac{42\pi^2}{5}\left(\frac{k_BT}{\epsilon_f}\right)^2 + \frac{31\pi^4}{7} \left(\frac{k_B T}{\epsilon_f}\right)^4 \right),	\\
	S 				&=	-\frac{4\pi^2}{3}\frac{k_B}{e} \frac{k_B T}{\epsilon_f}\left( 1 - \frac{3\pi^2}{5} \left( \frac{k_B T}{\epsilon_f} \right)^2 \right),\\
	\kappa_{xx}		&=	\frac{k_B^2T\epsilon_f}{24\pi^3\hbar^2v_f f(\alpha)}\left[1+\frac{46\pi^2}{15} \left( \frac{k_B T}{\epsilon_f} \right)^2  \right],
\end{align}
keeping only the lowest correction to $S$ and $\kappa_{xx}$. When $k_BT \ll \epsilon_f$, the Wiedemann-Franz law holds.

\subsection{Electron-Electron Interactions}
If the concentration of charged impurities approaches zero, the Fermi energy will approach the Weyl node. In this limit, we no longer expect scattering off charged impurities to dominate the transport properties due to absence of charged impurities to scatter electrons. At neutrality (and assuming there are no impurities present to dope the system), one instead expects electron-electron interactions to dominate relaxation processes due to the weak screening of the Coulomb interaction near the Fermi point, similar to graphene.\cite{RevModPhys.83.407} This situation is in contrast to normal metals, {\it i.e.} Fermi liquids, where electron-electron interactions do not provide an efficient relaxation method even if they are strong.\cite{Ashcroft}  Near a Weyl point the only energy scale is the temperature, so the self energy is expected to be proportional to the temperature. \cite{PhysRevB.84.235126,PhysRevLett.108.046602} Thus, as pointed out in Ref.~[\onlinecite{PhysRevLett.108.046602}] and by BHB, the inverse scattering time for electron-electron interactions (when $0\leq\mu<k_BT$) is (up to log corrections),
\begin{equation}
\frac{1}{\tau_{e-e}}=2\mathrm{Im}\Sigma=\frac{1}{A}\alpha^2 T,
\label{eq:tau_interactions}
\end{equation}
where $A$ is a proportionality constant.  We note for $\mu>k_B T$, the self-energy is proportional to the energy of the quasi-particle, not the temperature.\cite{PhysRevB.84.235126} The relaxation time for electron-electron interactions can also be worked out explicitly from field theoretical methods, as was done in Ref.~[\onlinecite{Abrikosov}]. Using Eq.~\eqref{eq:tau_interactions} as the scattering time, and evaluating the integral in Eq.~\eqref{mathL}, we find

\begin{align}
	\sigma_{xx}		&=	\frac{Ae^2k_B^2 T}{18\alpha^2 \hbar^3 v_f} \left[ 1+\frac{3}{\pi^2} \left( \frac{\epsilon_f}{k_B T} \right)^2 \right],\\
	L^{12}_{xx}		&=	-\frac{Aek_B^2  \epsilon_f}{9\alpha^2 \hbar^3 v_f},\\
	L^{22}_{xx}		&=	\frac{Ak_B^4T^2}{90\alpha^2\hbar^3 v_f} \left[ 7\pi^2 + 5 \left( \frac{\epsilon_f}{k_B T} \right)^2  \right],\\
	S 				&=	-2\frac{k_B}{e}\frac{\epsilon_f}{k_BT} \left[ 1 - \frac{3}{\pi^2} \left( \frac{\epsilon_f}{k_B T} \right)^2 \right],\\
	\kappa_{xx}		&=	\frac{A k_B^4 T^2}{90\alpha^2 \hbar^3 v_f} \left( 7\pi^2  -15 \left( \frac{\epsilon_f}{k_BT} \right)^2 \right).
\end{align}
Again we keep only the lowest correction to $S$ and $\kappa_{xx}$. When interactions dominate transport, the Wiedemann-Franz law is modified and we have (for $\epsilon_f\ll k_BT$)
\begin{equation}
\frac{\kappa_{xx}}{\sigma_{xx}T}=\frac{7\pi^2 k_B^2}{5e^2},
\end{equation}
which amounts to a change in the numerical prefactor. A similar modification of the Wiedemann-Franz law occurs in graphene, where it has recently been experimentally verified.\cite{PhysRevX.3.041008} In Fermi liquids, however, interactions do not change the value of the prefactor.\cite{PhysRevLett.59.477} We also note that the exact ratio of thermal conductivity and electrical conductivity is hard to experimentally determine due to the contact resistance.

\subsection{Short-Range Disorder}
We finish with the case of short range disorder, which is less realistic for Weyl and Dirac semimetals because the relatively poor screening of charged impurities (the most likely type) will lead to longer-range potentials.  Nevertheless, it is useful to investigate the predictions for the thermal properties in this case for purposes of comparison.  We ignore rare region effects,\cite{PhysRevB.89.245110} which give rise to a exponentially small density of states at the Weyl node. Again, the scattering time was calculated in the first Born approximation, which is valid for weak disorder, by BHB using the following short range potential for disorder
\begin{equation}
V(r)=\sum_a u_0 \delta(r-r_a),
\end{equation}
where $r_a$ label the impurity positions and $u_0$ is the strength of the zero-range impurity potential. BHB found the scattering time to be given by
\begin{equation}
\frac{1}{\tau_{\rm disorder}}=2\pi \gamma g(\epsilon),	\label{taudisorder}
\end{equation}
where $\gamma=u_0^2n_d$ and $n_d$ is the concentration of impurities. Ref.~[\onlinecite{PhysRevB.89.014205}] noted that this was the state lifetime and including vertex corrections introduces a factor of $\frac{3}{2}$ between the state lifetime and the transport time, i.e. $\tau_{tr}=\frac{3}{2}\tau$. In this section, we use the transport time to obtain the thermoelectric coefficients.  One expects disorder to dominate near the Weyl point, at high temperatures, in which case the energy $\epsilon=k_B T$. Since the total scattering rate goes as $1/\tau_{\rm total}=1/\tau_{e-e}+1/\tau_{\rm disorder}$, the condition for disorder to dominate the relaxation process is $1/\tau_{\rm disorder} \gg 1/\tau_{e-e}$, or $\frac{\tau_{\rm disorder}}{\tau_{e-e}}=\frac{\pi \hbar^3 v_f^3 \alpha^2}{A n_d u_0^2 k_B} \frac{1}{k_B T} \ll1$.

Using Eq.~\eqref{taudisorder} (and taking into account vertex corrections) for the relaxation time, and evaluating the integral in Eq.~\eqref{mathL}, we find
\begin{align}
	\sigma_{xx}=\frac{e^2v_f^2}{2\gamma h},\\ 
	L^{12}_{xx}=0,\\
	L^{22}_{xx}=\frac{\pi^2 v_f^2 k_B}{6\gamma h}k_B T,\\
	S 				=0,\\
	\kappa_{xx}			=\frac{\pi^2 v_f^2 k_B}{6\gamma h}k_B T.
\end{align}
Note that $L^{12}_{xx}$ and the thermopower $S$ are zero due to the integrand in Eq.~\eqref{mathL} being odd. We stress that this is a consequnce of using the first Born approximation, not a physical result. In Appendix A, we use the results of Ref.~[\onlinecite{PhysRevB.89.014205}] from the self-consistent Born approximation to estimate the lowest order corrections to the scattering time and find 
\begin{equation}
\tau_{\rm disorder}=\frac{3}{2}\frac{1}{2\pi\gamma g(\epsilon)} \left(1+\frac{5}{16\pi^2}\frac{\gamma^2}{\hbar^6v_f^6}\epsilon^2\right).
\end{equation}
Evaluating the integral in Eq.~\eqref{mathL} for these corrections, we find
\begin{align}
	L^{12}_{xx}		=-\frac{5}{96\pi}\frac{e\gamma k_B}{v_f^4\hbar^7}k_B T \epsilon_f, \\
	S 				=-\frac{5}{24}\frac{\gamma^2 }{v_f^6\hbar^6}\frac{k_B}{e} k_BT \epsilon_f
\end{align}
to lowest order in disorder strength. As in the case of relaxation due to charged impurities, the Wiedemann-Franz law holds for relaxation due to weak short-range disorder. We note that for strong disorder, there is a crossover to diffusive behavior. \cite{PhysRevB.89.054202,PhysRevLett.113.026602} The transition to diffusive behavior is beyond the scope of this paper and cannot be captured within the first Born approximation.

\begin{table*}
\renewcommand{\arraystretch}{1.8}
\begin{tabular}{|l|c|c|c|c|c|}
\hline
Relaxation Method & ~~~~~$\sigma_{xx}$~~~~~ &~~~~~ $L^{12}_{xx}$~~~~~&~~~~~ $L^{22}_{xx}$ ~~~~~& ~~~~~$S$ ~~~~~& ~~~~~$\kappa_{xx}$~~~~~\tabularnewline
\hline
Charged Impurities & ~~~~~$\frac{e^2 \epsilon_f}{8\pi^5 v_f \hbar^2 f(\alpha )}$~~~~~ & ~~~~~ $-\frac{ek_B^2 T}{6\pi^3 \hbar^2 v_f f(\alpha)}$~~~~~& ~~~~~$ \frac{k_B^2 T\epsilon_f}{24\pi^3\hbar^2v_f f(\alpha)}  $~~~~~& ~~~~~$-\frac{4\pi^2}{3}\frac{k_B}{e}(\frac{k_BT}{\epsilon_f} ) $ ~~~~~& ~~~~~$\frac{k_B^2T\epsilon_f}{24\pi^3\hbar^2v_f f(\alpha)}$~~~~~\tabularnewline
\hline
Electron-Electron Interactions & ~~~~~$\frac{Ae^2k_B^2 T}{18\alpha^2 \hbar^3 v_f}$ ~~~~~& ~~~~~$-\frac{Aek_B^2  \epsilon_f}{9\alpha^2 \hbar^3 v_f}$~~~~~ &~~~~~ $\frac{7\pi^2 A k_B^4 T^2}{90\alpha^2 \hbar^3 v_f} $~~~~~ & ~~~~~$-2\frac{k_B}{e}(\frac{\epsilon_f}{k_BT})$~~~~~ & ~~~~~$\frac{7\pi^2 A k_B^4 T^2}{90\alpha^2 \hbar^3 v_f} $~~~~~\tabularnewline
\hline
Short-Range Disorder   & ~~~~~$\frac{e^2v_f^2}{2\gamma h}$~~~~~ &~~~~~ $-\frac{5}{96\pi}\frac{e\gamma k_B}{v_f^4\hbar^7}k_B T \epsilon_f$~~~~~& ~~~~~$\frac{\pi v_f^2}{6\gamma h}k_B T$~~~~~ &~~~~~ $-\frac{5}{24}\frac{\gamma^2 }{v_f^6\hbar^6}\frac{k_B}{e} k_BT \epsilon_f$ ~~~~~& ~~~~~$\frac{\pi v_f^2}{6\gamma h}k_B T$~~~~~\tabularnewline
\hline
\end{tabular}
\caption{Transport coefficients for various relaxation methods, in the absence of magnetic and electric fields. All variables and regimes of validity are given in the main text. The electrical conductivities $\sigma_{xx}$ were, up to numerical factors, previously obtained in Ref.~[\onlinecite{PhysRevB.84.235126}].}
\label{Nofields}
\end{table*}

\subsection{Comparison of Results}
Before studying the effect of magnetic and electric fields on the thermal transport, we briefly compare our zero-field results to each other and contrast our results with other phases of matter that do not possess the three-dimensional linear dispersion. First, the temperature dependence of the thermal conductivity and thermopower can be the same with respect to different scattering processes (for $\frac{k_BT}{\epsilon_f}\gg1$). As an explicit example of this, we see that the thermal conductivity due to scattering off charged impurities has the same linear temperature dependence as scattering off short-range disorder. However, for scattering rates that are independent of the temperature, most common band structures, including quadratic band structures, will have the same linear temperature dependence, so this feature does not serve as an identifier of a Weyl or Dirac semimetal.\cite{Ashcroft} Experimentally investigating the transport coefficients as a function of the Fermi energy (perhaps through gating a sample), would provide a clearer experimental signature of the Dirac or Weyl semimetal, compared to measuring the temperature dependence. Measuring just the longitudinal thermoelectric coefficients, would not allow one to distinguish between a Weyl or Dirac semimetal.  We note that the quadratic temperature dependence of thermal conductivity is interesting when electron-electron interaction dominates relaxation processes since the scattering time then depends on the temperature. One experimental feature to look for in a three-dimensional system with linear electronic dispersion would be a crossover in the temperature dependence of the thermal conductivity from quadratic to linear upon adding/removing charged impurities (dopants) or disorder.

Of course, in real materials transport will be determined by a mix of all scattering processes, and we expect Matthiessen's rule, $1/\tau_{\rm total}=\sum_i 1/\tau_i$ where $\tau_i$ are the rates from different scattering processes, to apply,\cite{Ashcroft} assuming these scattering processes can be treated as independent. We now discuss the regimes of validity for each scattering process. We begin with a clean Weyl semimetal with the Fermi level at the Weyl node, in which case the transport is determined by electron-electron interactions. We then imagine adding charged impurities, which change the Fermi energy. We can find the condition for scattering off charged impurities to dominate transport due to electron-electron interactions by comparing inverse scattering rates (assuming a Fermi energy away from the nodal point): $\tau_{\rm screened}/\tau_{e-e}\sim \frac{\epsilon_f^2 T}{n_i} \ll 1$, which occurs at low temperatures.  We now imagine adding short-range (uncharged) disorder to the system.  Again comparing inverse scattering rates, scattering from short range impurities will dominate transport when $\tau_{\rm disorder}/\tau_{e-e} \ll 1$ and when $\tau_{\rm disorder}/\tau_{\rm screened} \ll 1$.  This requires high enough temperatures to neglect electron-electron scattering, but also the condition $\frac{\hbar^6 v_f^6 n_i}{u_0^2 n_d \epsilon_f^4} f(\alpha) \ll 1 $. The results for the three regimes are summarized in Table~\ref{Nofields}.

\section{Thermoelectric Transport Coefficients in Electric and Magnetic Fields}\label{sec:TTCEB}
In the presence of magnetic and electric fields, transport properties acquire modifications due to the Berry phase.\cite{RevModPhys.82.1959} A modified Boltzmann equation, which takes into account the chiral anomaly has been developed in several recent works.\cite{PhysRevD.87.085016,PhysRevLett.110.262301,PhysRevLett.109.181602,PhysRevB.88.104412} The chiral anomaly can be derived in several ways, including from the four-dimensional quantum Hall effect. \cite{2001Sci...294..823Z}  It can be treated mathematically by explicitly inserting a space-time dependent $\theta$-term in the action that couples electric and magnetic fields.\cite{PhysRevB.86.115133} The Boltzmann equation for a single Weyl node (for a given chirality) is given by 
\begin{equation}
	\left( \frac{\partial}{\partial t} + \dot{\vec{r}}^\chi \cdot \vec{\nabla}_r + \dot{\vec{p}}^\chi \cdot \vec{\nabla}_p \right)	f^\chi	= \mathrm{I}_{coll}^{\chi}
\end{equation}
where the modified semiclassical equations of motion are\cite{PhysRevLett.109.181602} 
\begin{align}
	\dot{\vec{r}}^\chi	&=	\left(1+\frac{e}{c}\vec{B}\cdot\vec{\Omega}^{\chi}\right)^{-1} \left[ \vec{v} + e\vec{E} \times \vec{\Omega}^\chi + \frac{e}{c} \left(\vec{\Omega}^\chi \cdot \vec{v}\right) \vec{B} \right],	\label{rdot}\\
	\dot{\vec{p}}^\chi	&=	\left(1+\frac{e}{c}\vec{B}\cdot\vec{\Omega}^{\chi}\right)^{-1} \left[ e\vec{E} + \frac{e}{c}\vec{v}\times \vec{B} + \frac{e^2}{c} \left( \vec{E} \cdot \vec{B} \right) \vec{\Omega}^\chi \right],
\end{align}
where $\vec{B}$ is the magnetic field and c is the speed of light. Plugging the modified semiclassical equations of motion (which include the Berry phases associated with the Weyl points) in the Boltzmann equation and using the relaxation time approximation for the collision integral, $I^\chi_{coll}=-\frac{f^\chi-f_{eq}}{\tau}-\frac{f^{\chi}-f^{-\chi}}{\tau_s}$, where $\tau_s$ is inter-node scattering time and $\tau$ is the intra-node scattering time as before, gives
\begin{widetext}
\begin{align}
\frac{\partial{f^{\chi}}}{\partial t}+\left(1+\frac{e}{c}\vec{B}\cdot\vec{\Omega}^{\chi}\right)^{-1}\left\{ \left(\vec{v}+e\vec{E}\times\vec{\Omega}^{\chi}+\frac{e}{c}\left(\vec{\Omega}^{\chi}\cdot\vec{v}\right)\vec{B}\right)\cdot\vec{\nabla}_r f^{\chi}+
\left(e\vec{E}+\frac{e}{c}\vec{v}\times\vec{B}+\frac{e^2}{c}\left(\vec{E}\cdot\vec{B}\right)\vec{\Omega}^{\chi}\right)\cdot\vec{\nabla}_p f^{\chi}\right\} \nonumber \\
=-\frac{f^\chi-f_{eq}}{\tau}-\frac{f^{\chi}-f^{-\chi}}{\tau_s}.
\label{FullB}
\end{align}
For simplicity, we neglect internode scattering, {\it i.e.} $\tau_s\rightarrow \infty$, and treat the intranode scattering time $\tau$ as a phenomenological parameter. We limit ourselves to two Weyl nodes, but it is a straightforward generalization to more nodes, as the nodes contribute additively. The additive effect when a magnetic field is present, is the cancellation of some terms due to opposite chirality. The magnetic field modifies the distance and direction between Weyl nodes, and the Berry curvature is proportional to\cite{PhysRevB.89.195137}
\begin{equation}
\vec{\Omega}^{\chi}\propto\chi\frac{\hat{p}}{|\vec{p}-g\chi \vec{B}|},
\end{equation}
where $g$ is the Land\'{e} g-factor and $\vec{p}=\hbar \vec k$ is the quasiparticle momentum. This approach is valid when the Fermi energy lies away from the Weyl node, $\mu \gg k_B T$,  and $\mu>\hbar \omega_c$, where $\omega_c=\frac{eB v_f^2}{c\mu}$ is the cyclotron frequency. \cite{PhysRevB.89.195137} In the presence of a magnetic field at finite charge density, a Dirac semimetal becomes a Weyl semimetal, \cite{PhysRevB.88.165105} thus our approach applies to both Weyl and Dirac semimetals, provided one can ignore inter-node scattering, as we discussed earlier. In our approach we treat the scattering time as a phenomenological parameter, {\it i.e.} we set the quasi-particle energy that appears in the scattering time equal to the chemical potential. Treating the scattering time as independent of quasi-particle energy is a valid assumption given that this semiclassical treatment is only valid at large chemical potential and the scattering times considered earlier decrease when the chemical potential increases. Stated more explicitly, whenever the scattering time appears, the integrand that determines the transport coefficients is centered around $\mu$ with its width proportional to $k_BT$ (due to the term with the derivative of $f_{eq}$ with respect to energy). \cite{Ashcroft} The scattering times we considered, which decrease rapidly with $\mu$, do not change appreciably over this width given that our approach is only valid when $\mu\gg k_BT$. Thus, we are able to treat the scattering time phenomenologically. We also note that for the scattering times considered earlier when $\mu>T$ are independent of temperature.  We stress in our model that the distance between the Weyl nodes is determined by the magnetic field. If the distance between the Weyl nodes is determined by something other than the magnetic field (which happens in stacked layers of three dimensional topological insulators and normal insulators\cite{PhysRevLett.107.127205} for example), one expects the thermoelectric coefficients to have a different dependence on the magnetic field. We now turn to a discussion of a few important special cases of Eq.~\eqref{FullB} where the electric and magnetic fields are applied along certain high-symmetry directions.

\begin{figure}
\subfloat[][$\vec{\nabla}T=\nabla T \hat{x},\vec{B}=B \hat {x}$]{
 \includegraphics[width=.5\linewidth]{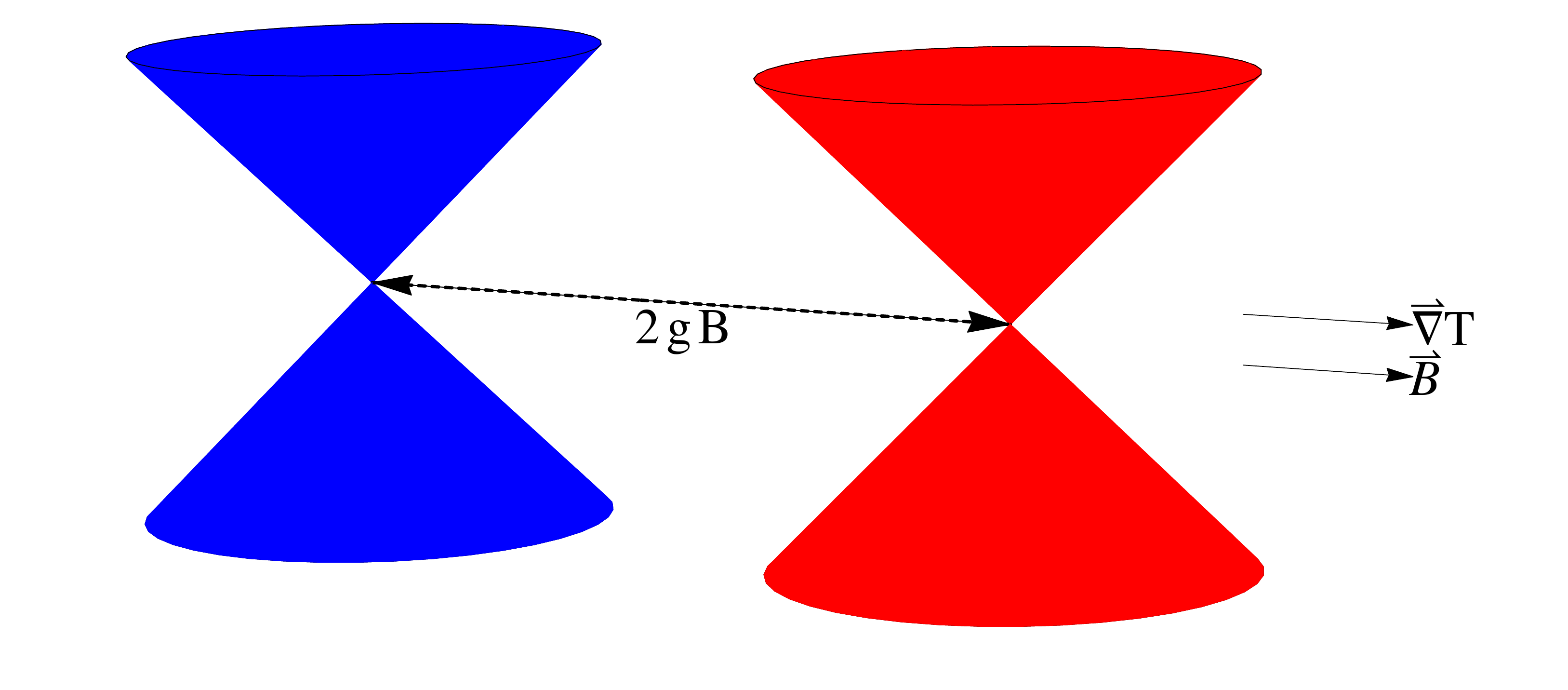}
}
\subfloat[][$\vec{\nabla}T=\nabla T \hat{x},\vec{B}=B \hat {z}$]{
 \includegraphics[width=.5\linewidth]{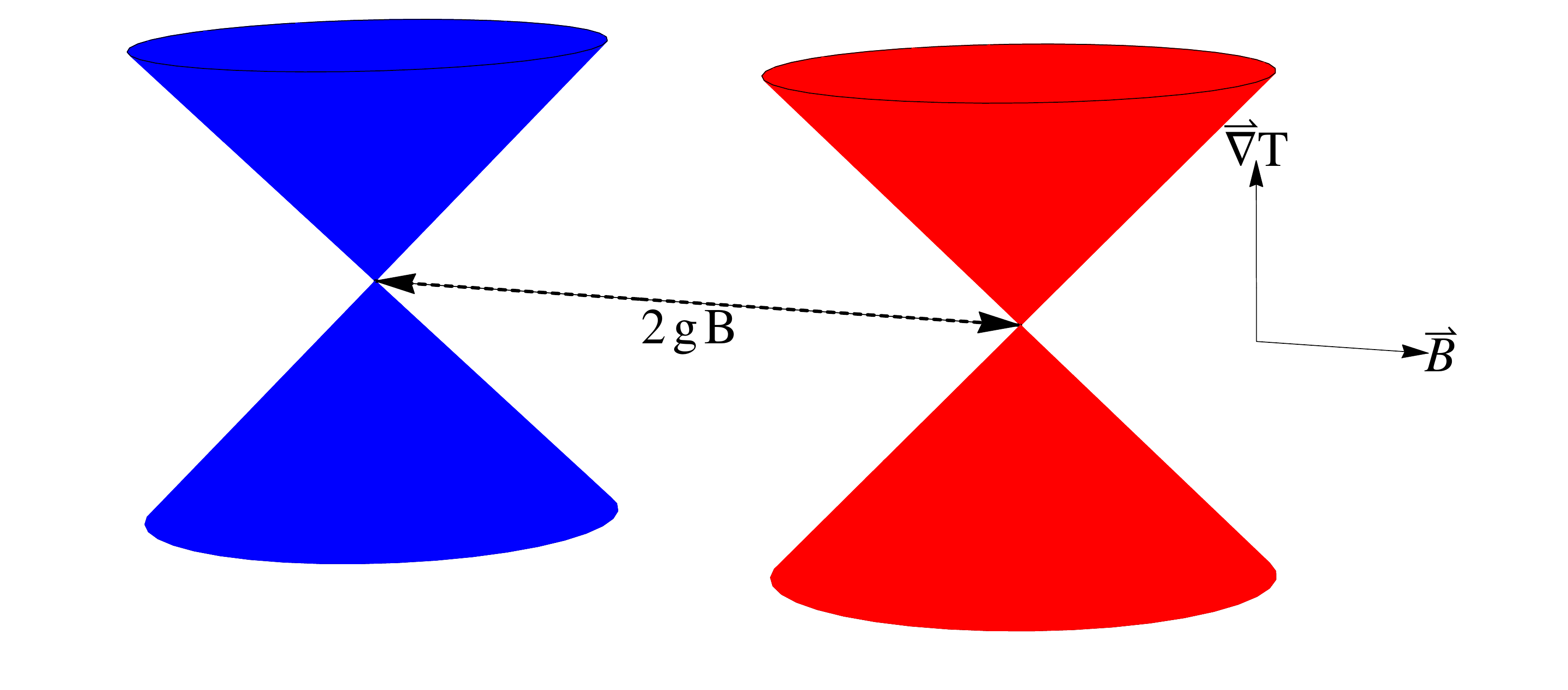}
}
\caption{(color online) Illustration of two Weyl cones separated by $2gB$ in momentum space with various thermal gradients and magnetic field orientations. The different color of the two nodes highlights the fact that each node has a definite chirality.}
\label{FiguresNOE}
\end{figure}

\begin{figure}
\subfloat[][$\vec{\nabla}T=\nabla T \hat{x},\vec{B}=B \hat {x},\vec{E}=E\hat{z}$]{
 \includegraphics[width=.33\linewidth,height=.18\textwidth]{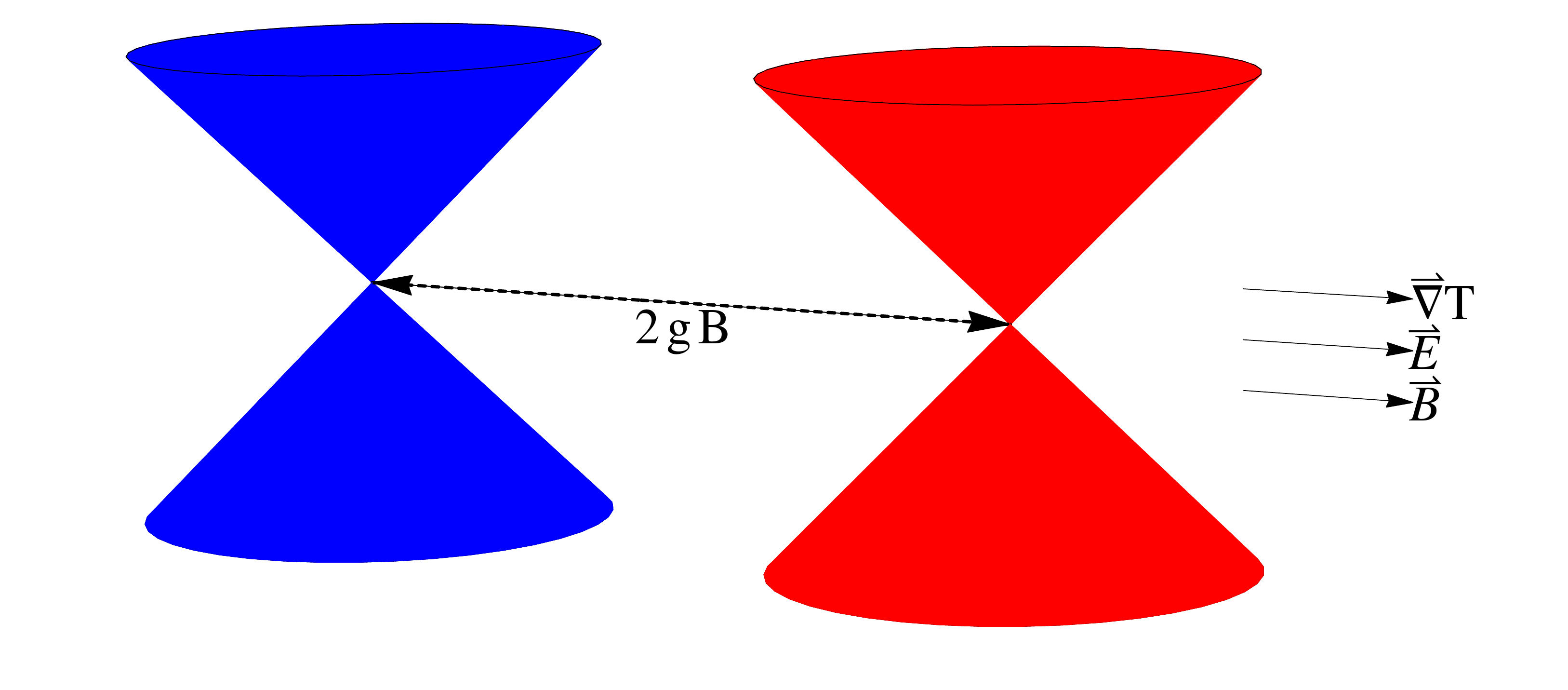}
\label{ALL_PAR}
\newline
}
\subfloat[][$\vec{\nabla}T=\nabla T \hat{x},\vec{B}=B \hat {z},\vec{E}=E\hat{z}$]{
 \includegraphics[width=.33\linewidth,height=.18\textwidth]{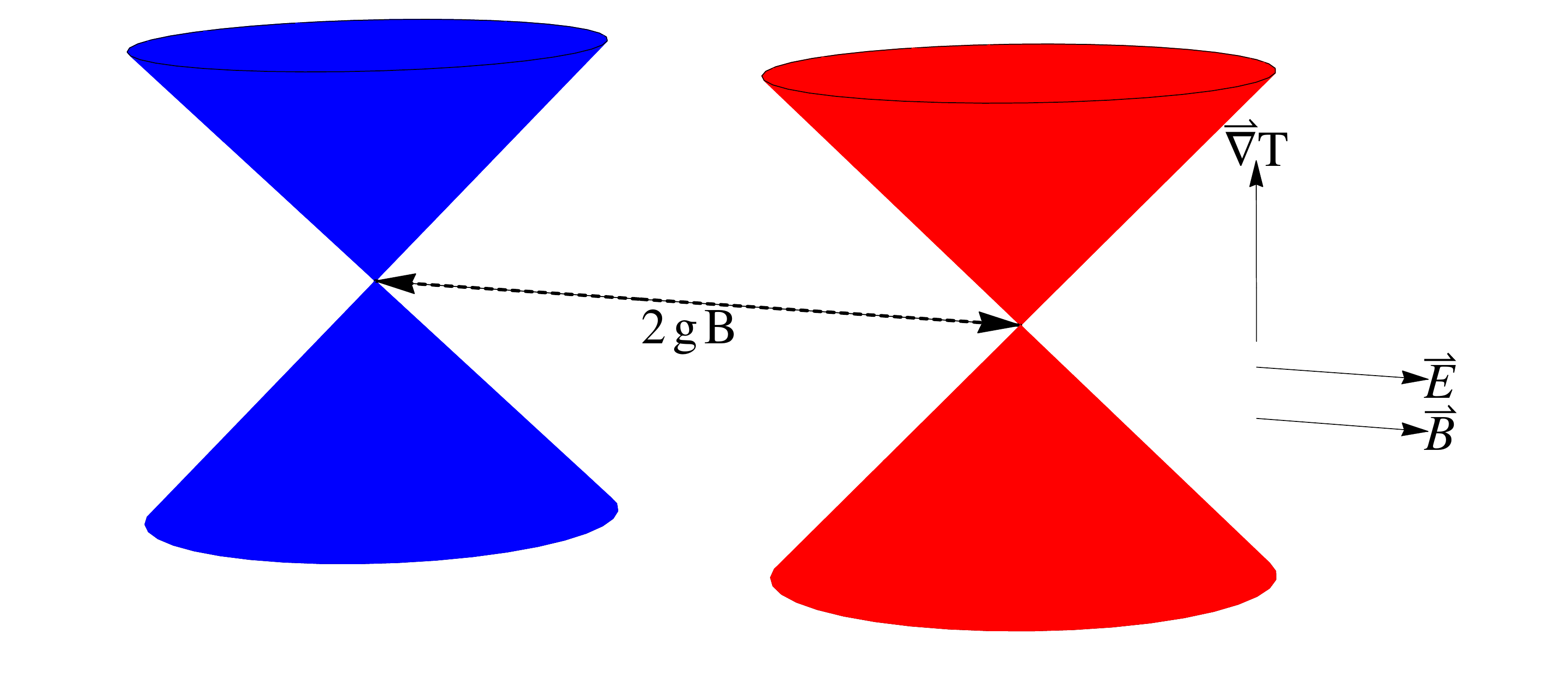}
\label{EB_PAR}
}
\subfloat[][$\vec{\nabla}T=\nabla T \hat{x},\vec{B}=B \hat {y},\vec{E}=E\hat{z}$]{
 \includegraphics[width=.33\linewidth,height=.18\textwidth]{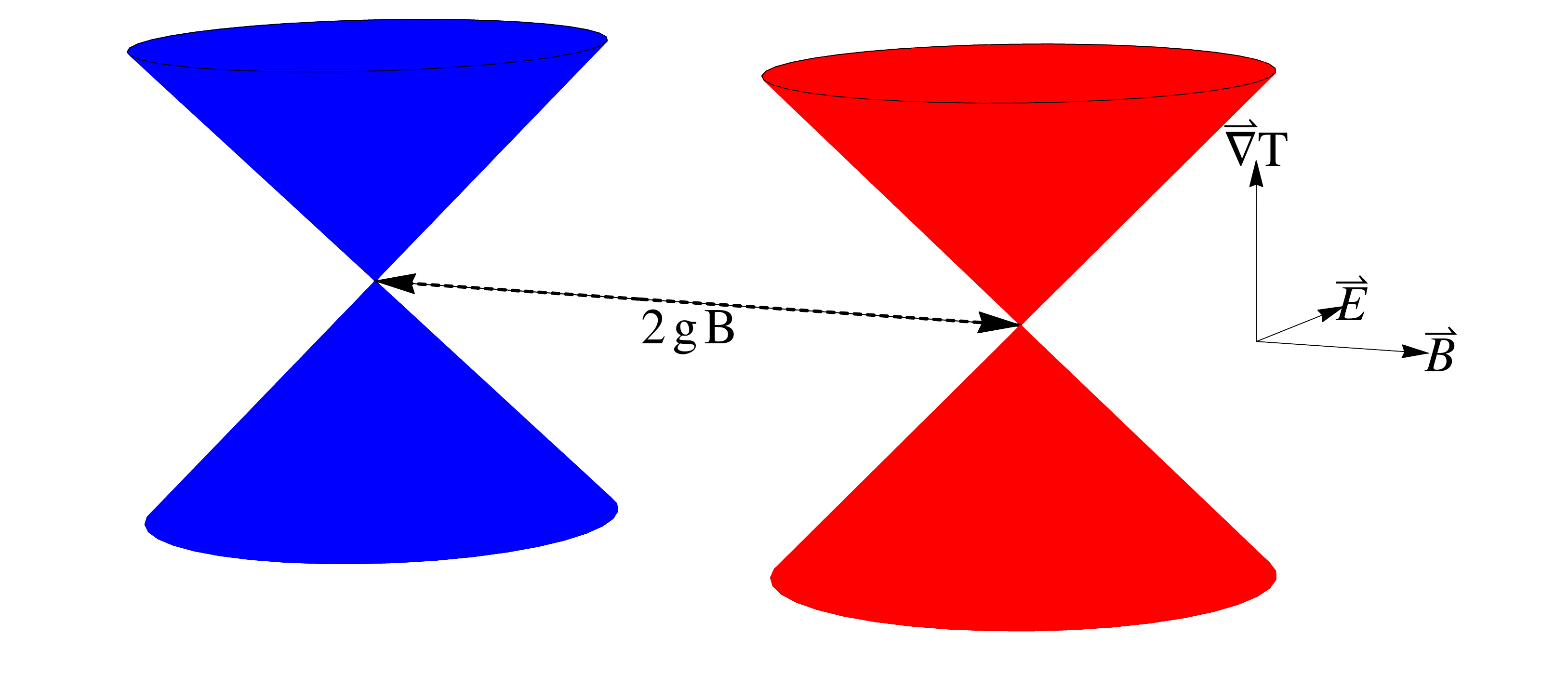}
\label{ALL_PERP}
}
\caption{(color online) Illustration of two Weyl cones separated by $2gB$ in momentum space with various thermal gradients and electric and magnetic field orientations.}
\label{FiguresE}
\end{figure}

\subsection{$\vec{\nabla}T=\nabla T \hat{x},\vec{B}=B \hat {x},\vec{E}=0$}\label{sec:TBPAR}

We now consider the thermoelectric coefficients when the temperature gradient is parallel to the magnetic field and the electric field is zero, as depicted in Fig.~\ref{FiguresNOE}(a). In this case, the Boltzmann equation for a given chirality reads
\begin{align}
\frac{\partial{f^{\chi}}}{\partial t}+\left(1+\frac{e}{c}\vec{B}\cdot\vec{\Omega}^{\chi}\right)^{-1}\left\{\left(\vec{v}+\frac{e}{c}\left(\vec{\Omega}^{\chi}\cdot\vec{v}\right)\vec{B}\right)\cdot \vec{\nabla}_r f^{\chi}+
\left(\frac{e}{c}\vec{v}\times\vec{B}\right)\cdot\vec{\nabla}_p f^{\chi}\right\}=-\frac{f^\chi-f_{eq}}{\tau}.
\label{f_chi_t}
\end{align}
Assuming steady state and linear response, the equation can be written
\begin{align}
	\left\{ \frac{1}{\tau} + \left(1+\frac{e}{c}\vec{B}\cdot\vec{\Omega}^{\chi}\right)^{-1} \frac{eB}{c} \left( v_z \frac{\partial}{\partial p_y} - v_y \frac{\partial}{\partial p_z} \right) \right\} f^\chi	&=	\frac{1}{\tau} f_{eq} - \left(1+\frac{e}{c}\vec{B}\cdot\vec{\Omega}^{\chi}\right)^{-1} \left\{ v_x + \frac{e}{c} \left( \vec{\Omega}^\chi \cdot \vec{v} \right) \right\} \nabla T \frac{\partial f_{eq}}{\partial T}	\label{f_chi_t_linear}.
\end{align}
Following Ref. [\onlinecite{PhysRevB.89.195137}], we assume the following Ansatz for the distribution function
\begin{equation}
	f^\chi	=	f_{eq} - \left( 1+ \frac{e}{c} \vec{B} \cdot \vec{\Omega}^\chi \right)^{-1} \tau\left( v_x \nabla T \frac{\epsilon-\mu}{T} \left( -\frac{\partial f_{eq}}{\partial \epsilon}\right) +\frac{e}{c} \left(\vec{\Omega}^{\chi}\cdot \vec{v}\right) B \nabla T \frac{\epsilon-\mu}{T}\left( -\frac{\partial f_{eq}}{\partial \epsilon}\right)\right) + \left( -\frac{\partial f_{eq}}{\partial \epsilon} \right) \vec{v} \cdot \vec{\Lambda}^\chi,
\end{equation}
based off the structure of Eq.~\eqref{f_chi_t_linear} and using $\frac{\partial f_{eq}}{\partial T} = \frac{\epsilon-\mu}{T} \left( - \frac{\partial f_{eq}}{\partial \epsilon}\right)$. Here $\vec{\Lambda}^\chi$ represents a correction term due to the magnetic field. Using the condition that the Boltzmann equation must hold for all values of $\vec{v}$, we find the corrections $\Lambda^\chi_x=0$ and
\begin{align}
	\Lambda_y^\chi	&=	\tau \frac{eB}{c} \frac{\nabla T\frac{\epsilon-\mu}{T}}{\left( 1 + \frac{e}{c} \vec{B} \cdot \vec{\Omega}^\chi \right)^2 + \omega_c^2 \tau^2} \left[ \Omega_z^\chi \omega_c \tau - \left( 1+ \frac{e}{c} \vec{B} \cdot \vec{\Omega}^\chi \right)^{-1} \Omega_y^\chi \omega_c^2 \tau^2 \right], \label{arrow_y}	\\
	\Lambda_z^\chi	&=	\tau \frac{eB}{c} \frac{\nabla T\frac{\epsilon-\mu}{T}}{\left( 1 + \frac{e}{c} \vec{B} \cdot \vec{\Omega}^\chi \right)^2 + \omega_c^2 \tau^2} \left[ -\Omega_y^\chi \omega_c \tau - \left( 1+ \frac{e}{c} \vec{B} \cdot \vec{\Omega}^\chi \right)^{-1} \Omega_z^\chi \omega_c^2 \tau^2 \right]. \label{arrow_z}
\end{align}

We define the electrical current as in Eq. \eqref{electric_current} and the heat current as in Eq. \eqref{Thermal_current}, where $\vec{v}$ must now be replaced by $\left( 1+\frac{e}{c}\vec{B}\cdot\vec{\Omega}^\chi\right) \dot{\vec{r}}$, where $\dot{\vec{r}}$ is the modified semiclassical velocity given in Eq. \eqref{rdot}, in order to take the effects of the chiral anomaly into account. Specializing to transport along the direction of the temperature gradient, we find
\begin{align}
	J^{\chi}_x	=-e\int\frac{\mathrm{d}^3p}{(2\pi)^3} \left( 1+\frac{e}{c}\vec{B}\cdot\vec{\Omega}^\chi\right) \dot{\vec{r}} f^{\chi}	&= e\int \frac{\mathrm{d}^3p}{(2\pi)^3} \left( 1+ \frac{e}{c} \vec{B} \cdot \vec{\Omega}^\chi \right)^{-1} \tau \left( v_x  +\frac{e}{c} \left(\vec{\Omega}^{\chi}\cdot \vec{v}\right) B \right)^2 \nabla T \frac{\epsilon-\mu}{T}\left( -\frac{\partial f_{eq}}{\partial \epsilon}\right), \\
	J_{q,x}^{\chi}	= \int\frac{\mathrm{d}^3k}{(2\pi)^3}(\epsilon-\mu) \left( 1+\frac{e}{c}\vec{B}\cdot\vec{\Omega}^\chi\right) \dot{\vec{r}}  f^{\chi} &=-\int \frac{\mathrm{d}^3k}{(2\pi)^3} \left( 1+ \frac{e}{c} \vec{B} \cdot \vec{\Omega}^\chi \right)^{-1} \tau \left( v_x  +\frac{e}{c} \left(\vec{\Omega}^{\chi}\cdot \vec{v}\right) B \right)^2 \nabla T \frac{\left( \epsilon-\mu \right)^2}{T}\left( -\frac{\partial f_{eq}}{\partial \epsilon}\right),
\end{align}
which allows us to write down the expressions for $L^{12}_{xx}$ and $L^{22}_{xx}$. Discarding the term linear in $\vec{\Omega}^\chi$ (and hence chirality) since it will vanish when we sum over both nodes, noting $\vec{\Omega}^\chi \propto \hat{p} \propto \hat{v}$, and using $v_x^2=v_f^2/3$, we find
\begin{align}
	L^{12}_{xx}		=	-e&\int \frac{\mathrm{d}^3p}{(2\pi)^3} \left( 1+ \frac{e}{c} \vec{B} \cdot \vec{\Omega}^\chi \right)^{-1} \tau \frac{v_f^2}{3} \left( 1 + \left(\frac{eB}{c} \right)^2 \left| \vec{\Omega}^\chi \right|^2 \right) \frac{\epsilon-\mu}{T}\left( -\frac{\partial f_{eq}}{\partial \epsilon}\right), \\
	L^{22}_{xx}		=&\int \frac{\mathrm{d}^3p}{(2\pi)^3} \left( 1+ \frac{e}{c} \vec{B} \cdot \vec{\Omega}^\chi \right)^{-1} \tau \frac{v_f^2}{3} \left( 1 + \left(\frac{eB}{c} \right)^2 \left| \vec{\Omega}^\chi \right|^2 \right) \frac{\left( \epsilon-\mu\right)^2}{T}\left( -\frac{\partial f_{eq}}{\partial \epsilon}\right).
\end{align}
To calculate the electronic thermal conductivity, we also need the electrical conductivity $\sigma_{xx}$, or in other words $L^{11}_{xx}$. For the case of perpendicular electric and magnetic fields, Ref. [\onlinecite{PhysRevB.89.195137}] found it to be
\begin{equation}
\sigma_{xx}=e^2\int\frac{\mathrm{d}^3p}{(2\pi)^3}(1+\frac{e}{c}\vec{B}\cdot\Omega^{\chi})^{-1}\tau\left(-\frac{\partial f_{eq}}{\partial \epsilon}\right) \left(v_x+\frac{e}{c}B\vec{v}\cdot\vec{\Omega}^{\chi}\right)^2=e^2\int\frac{\mathrm{d}^3p}{(2\pi)^3}(1+\frac{e}{c}\vec{B}\cdot\Omega^{\chi})^{-1}\tau \left(-\frac{\partial f_{eq}}{\partial \epsilon}\right)\frac{|\vec{v}|^2}{3} \left(1+\left(\frac{eB}{c}\right)^2|\vec{\Omega}^{\chi}|^2\right).
\label{sigma_EB}
\end{equation}
by a similar calculation. This can be explicitly seen by setting the temperature gradient to zero in Eq.~\eqref{allparr}. Assuming a smoothly varying Berry phase, we can estimate this integral using a series expansion in $B$. There is a linear term proportional to $\Omega_x$, which will vanish in the angular momentum integration, leaving the electrical conductivity for the two Weyl nodes as
\begin{equation}
\sigma_{xx}\approx2\sigma_{xx}(B=0)\left(1+\left(\frac{eB}{c}\right)^2\frac{v_f^4}{\mu^4}\right),
\end{equation}
so that the magnetic field leads to a positive contribution to the conductivity that goes as the square of the field.
Here we have used the fact that $\Omega\propto\frac{1}{\mu^2}$. Estimating the other transport coefficients in a similar matter, we can obtain the thermal conductivity, which is written as a tensor\cite{Ashcroft}
\begin{equation}
\kappa_{\alpha,\beta}=L^{22}_{\alpha,\beta}-L^{12}_{\alpha,\gamma}\sigma^{-1}_{\gamma,\rho}L^{21}_{\rho,\beta}.
\label{eq:kappa_tensor}
\end{equation}
 We find the longitudinal thermal conductivity to be
\begin{equation}
\kappa_{xx}(B) \approx 2\kappa_{xx}(B=0)\left(1+\left(\frac{eB}{c}\right)^2\frac{v_f^4}{\mu^4}\right),
\label{TB}
\end{equation}
which also has a similar additivity contribution going as the square of the magnetic field.  We note the magnetic field dependent correction is independent of the scattering time and temperature.

We briefly comment on the transverse thermal transport, $\kappa_{xy}$, when the temperature gradient and magnetic field are parallel. As discussed earlier, we expect a term proportional to the distance between Weyl nodes due to the Berry curvature, given by Eq.~\eqref{Thermal_Hall_1}. For a magnetic field in the $x$-direction, this term is zero due to the Levi-Civita symbol. As pointed out in Ref.~[\onlinecite{PhysRevB.89.195137}], the contribution from the $\vec{E}\cdot\vec{B}$ term to $\sigma_{xy}$ vanishes due to the cancellation of terms linear in chirality when both Weyl nodes are added together. Thus, we expect the anomalous contribution from the $\vec{\nabla} T\cdot\vec{B}$ term to vanish for $\kappa_{xy}$ given the structure of the Boltzmann equation which treats the nodes as independent. This can be seen explicitly from Eqs.~\eqref{arrow_y} and \eqref{arrow_z}. Thus, in the case of a parallel magnetic field and temperature gradient, $\kappa_{xy}=0$.

\subsection{$\vec{\nabla}T=\nabla T \hat{x},\vec{B}=B \hat {z},\vec{E}=0$}\label{sec:TBPREP}
We now consider the case where the magnetic field is perpendicular to the temperature gradient, while the electric field is still zero (see Fig.~\ref{FiguresNOE}(b)). The steady-state Boltzmann equation for a given chirality (after expanding out the cross product) reads
\begin{align}
\left(1+\frac{e}{c}\vec{B}\cdot\vec{\Omega}^{\chi}\right)^{-1}\left(v_x\nabla T \frac{\epsilon-\mu}{T}\left(-\frac{\partial f_{eq}}{\partial \epsilon}\right)+\frac{eB}{c}\left(v_y\frac{\partial}{\partial p_y}f^{\chi}-v_x\frac{\partial}{\partial p_x}f^{\chi}\right)\right)=-\frac{f^{\chi}-f_{eq}}{\tau}.
\label{BoltzmmanBprepTnoE}
\end{align}
 Following Ref. [\onlinecite{PhysRevB.89.195137}] and the previous section, we assume the following solution for the Boltzmann equation
\begin{equation}
f^{\chi}=f_{eq}-\left(1+\frac{e}{c}\vec{B}\cdot\Omega^{\chi}\right)^{-1}\tau v_x \nabla T \frac{\epsilon-\mu}{T}\left(-\frac{\partial f_{eq}}{\partial \epsilon}\right)+\left(-\frac{\partial f_{eq}}{\partial \epsilon}\right)\vec{v}\cdot \vec{\Lambda}^{\chi},
\end{equation}
After some algebra, we find
\begin{equation}
\Lambda_{x}^{\chi}=\tau\nabla T \frac{\epsilon-\mu}{T}\frac{\left(1+\frac{e}{c}\vec{B}\cdot\vec{\Omega^{\chi}}\right)^{-1}\omega_c^2\tau^2}{\left(1+\frac{e}{c}\vec{B}\cdot\vec{\Omega}^{\chi}\right)^{2}+\omega_c^2\tau^2},
\end{equation}
\begin{equation}
\Lambda_y^{\chi}=\tau \nabla T \frac{\epsilon - \mu}{T} \frac{  \omega_c \tau}{ \left( 1+ \frac{e}{c} \vec{B} \cdot \vec{\Omega}^\chi \right)^{2}+ \omega_c^2 \tau^2},
\end{equation}
and $\Lambda_z^\chi=0$. Proceeding as in the previous section, we have
\begin{align}
	L^{12}_{xx}	&=	-e\int \frac{\mathrm{d}^3p}{(2\pi)^3}\left(1+\frac{e}{c}\vec{B}\cdot\Omega^{\chi}\right)^{-1} \tau v_x \left( v_x + \frac{e}{c} \left( \vec{\Omega}^\chi \cdot \vec{v} \right) B \right) \nabla T \frac{\epsilon-\mu}{T} \left( 1 - \frac{\omega_c^2 \tau^2}{\left(1+\frac{e}{c}\vec{B}\cdot\vec{\Omega}^{\chi}\right)^2+\omega_c^2\tau^2}\right) \left( -\frac{\partial f_{eq}}{\partial \epsilon} \right),\\
	L^{22}_{xx}	&=	\int \frac{\mathrm{d}^3p}{(2\pi)^3}\left(1+\frac{e}{c}\vec{B}\cdot\Omega^{\chi}\right)^{-1} \tau v_x \left( v_x + \frac{e}{c} \left( \vec{\Omega}^\chi \cdot \vec{v} \right) B \right) \nabla T \frac{\left( \epsilon-\mu \right)^2}{T} \left( 1 - \frac{\omega_c^2 \tau^2}{\left(1+\frac{e}{c}\vec{B}\cdot\vec{\Omega}^{\chi}\right)^2+\omega_c^2\tau^2}\right) \left( -\frac{\partial f_{eq}}{\partial \epsilon} \right).
\end{align}
Again, we need the longitudinal electrical conductivity when electric and magnetic fields are perpendicular to find the thermal conductivity. This was also calculated in Ref.~[\onlinecite{PhysRevB.89.195137}], by solving the Boltzmann equation when the electric and magnetic field are perpendicular and found to be 
\begin{equation}
\sigma_{xx}=e^2\int\frac{\mathrm{d}^3p}{(2\pi)^3}\left(1+\frac{e}{c}\vec{B}\cdot\Omega^{\chi}\right)^{-1}\left(-\frac{\partial f_{eq}}{\partial \epsilon}\right)\tau v_x^2\frac{1}{1+\frac{\omega_c^2\tau^2}{(1+\frac{e}{c}\vec{B}\cdot\vec{\Omega}^{\chi})^{2}}}.
\end{equation}
To lowest order in magnetic field and summing over both chiralities, we can estimate the thermal conductivity as
\begin{equation}
\kappa_{xx}\approx 2 \kappa_{xx}(B=0)\left(1-\left(\omega_c^2\tau^2\right)\right),
\label{LtwotwonoEfield}
\end{equation}
where we see that the thermal conductivity has the same dependence on the magnetic field as the electrical conductivity with perpendicular electric and magnetic fields, \cite{PhysRevB.89.195137} {\em but the sign is opposite}, which leads to a decrease in the thermal conductivity. The magnetic field dependence of thermal conductivity for perpendicular magnetic field and temperature gradient depends on the specific form of scattering, whereas the magnetic field dependence of thermal conductivity when the two fields are parallel does not. We emphasize that the thermal conductivity for perpendicular (parallel) magnetic field and thermal gradient decreases (increases) upon increasing magnetic field. We also note that Eq.~\eqref{LtwotwonoEfield} is the typical response of a material without Berry curvature and as such, is not unique to Dirac or Weyl semimetals.

We now investigate the transverse thermal transport. The transverse transport coefficients are given by
\begin{equation}
\sigma_{xy}=e^2\int\frac{\mathrm{d}^3p}{(2\pi)^3} \left( 1+ \frac{e}{c} \vec{B} \cdot \vec{\Omega}^\chi \right)^{-1}v_y^2\tau \frac{ \omega_c \tau}{ \left( 1+ \frac{e}{c} \vec{B} \cdot \vec{\Omega}^\chi \right)^{2}+ \omega_c^2 \tau^2}+\frac{e^2}{\hbar}\int\frac{\mathrm{d}^3p}{(2\pi)^3}\Omega_z^{\chi}f_{eq},
\end{equation}
\begin{equation}
L^{12}_{xy}=e\int\frac{\mathrm{d}^3p}{(2\pi)^3} \left( 1+ \frac{e}{c} \vec{B} \cdot \vec{\Omega}^\chi \right)^{-1}v_y^2(\epsilon - \mu)\tau \frac{ \omega_c \tau}{ \left( 1+ \frac{e}{c} \vec{B} \cdot \vec{\Omega}^\chi \right)^{2}+ \omega_c^2 \tau^2}+\frac{1}{T}\frac{e}{\hbar}\int\frac{\mathrm{d}^3p}{(2\pi)^3}\Omega_z^{\chi}(\epsilon-\mu)f_{eq},
\end{equation}
\begin{equation}
L^{22}_{xy}=\int\frac{\mathrm{d}^3p}{(2\pi)^3} \left( 1+ \frac{e}{c} \vec{B} \cdot \vec{\Omega}^\chi \right)^{-1}v_y^2\frac{(\epsilon - \mu)^2}{T}\tau \frac{ \omega_c \tau}{ \left( 1+ \frac{e}{c} \vec{B} \cdot \vec{\Omega}^\chi \right)^{2}+ \omega_c^2 \tau^2}+\frac{1}{T}\frac{1}{\hbar}\int\frac{\mathrm{d}^3p}{(2\pi)^3}\Omega_z^{\chi}(\epsilon-\mu)^2f_{eq}.
\label{eq:L22_anomalous}
\end{equation}
The last term in Eq.\eqref{eq:L22_anomalous} leads to the anomalous thermal Hall effect discussed earlier. To lowest order in magnetic field, we find $\kappa_{xy}$ for the two nodes to be 
\begin{equation}
\kappa_{xy}(B) \approx \frac{2\pi}{3h}k_B^2T g B-2 \kappa_{xx}(B=0)\omega_c\tau,
\label{transBnoE}
\end{equation}
where we have used the fact that distance between Weyl nodes is $2gB$ in Eq.~\eqref{Thermal_Hall_1}. We have ignored a small correction (for $k_BT \ll \mu$) due to the tensorial structure of the thermal conductivity, Eq.\eqref{eq:kappa_tensor}, on the order of $\left(\frac{k_BT}{\mu}\right)^2$.  In this case, we see that the anomalous thermal Hall conductivity is non-zero, as opposed to the case when the magnetic field and temperature gradient are parallel. However both terms have the same dependence on magnetic field, and even the same temperature dependence when longitudinal transport is dominated by disorder or charged impurities. Thus, to investigate the dependence of the thermal conductivity on magnetic fields, one could vary the Fermi energy, either by gating or other means. The first term  in Eq.\eqref{transBnoE} does not depend on the Fermi energy, while the second will have some dependence on Fermi energy, the specific form of which depends on the relaxation mechanisms that dominate $\tau$. This feature could in principle be used to extract the first term, the anomalous thermal Hall effect, by identifying the contribution to the transverse thermal conductivity that does not depend on the value of the Fermi energy.

\subsection{$\vec{\nabla}T=\nabla T \hat{x},\vec{B}=B \hat {x},\vec{E}=E\hat{x}$} \label{sec:ALLPAR}
The most interesting case is when both electric and magnetic fields are present, and they are parallel to each other (see Fig.~\ref{FiguresE}(a)), giving rise to the chiral anomaly and its novel effects. We first consider the case when the temperature gradient is parallel to the fields, in which case the Boltzmann equation in the relaxation time approximation becomes

\begin{equation}
	\left( 1+ \frac{e}{c} \vec{B} \cdot \vec{\Omega}^\chi \right)^{-1} \left[ \left( v_x + \frac{e}{c} \left( \vec{\Omega}^\chi \cdot \vec{v}\right) B \right) \nabla T \frac{\partial f^\chi}{\partial T} + eE\frac{\partial f^{\chi}}{\partial p_x} + \frac{e}{c} v_z B \frac{\partial f^{\chi}}{\partial p_y} - \frac{e}{c} v_y B \frac{\partial f^{\chi}}{\partial p_z} + \frac{e^2}{c} \left( \vec{E} \cdot \vec{B} \right) \vec{\Omega}^\chi \cdot \vec{\nabla}_p f^{\chi} \right]	= -\frac{f^{\chi}-f_{eq}}{\tau}.
\end{equation}
Following Ref. [\onlinecite{PhysRevB.89.195137}], we assume the following Ansatz for the distribution function
\begin{equation}
	f^\chi	=	f_{eq} - \left( 1+ \frac{e}{c} \vec{B} \cdot \vec{\Omega}^\chi \right)^{-1} \tau\left( eE \frac{\partial f_{eq}}{\partial p_x} + \frac{e^2}{c} (\vec{E}\cdot\vec{B}) \vec{\Omega}^{\chi} \cdot\vec{\nabla}_p f_{eq} + v_x \nabla T \frac{\partial f_{eq}}{\partial T} + \frac{e}{c} \left(\vec{\Omega}^{\chi}\cdot \vec{v}\right) B \nabla T \frac{\partial f_{eq}}{\partial T}\right) + \left( -\frac{\partial f_{eq}}{\partial \epsilon} \right) \vec{v} \cdot \vec{\Lambda}^\chi.
\label{allparr}
\end{equation}
We find that the corrections $\Lambda^\chi_x=0$ and
\begin{align}
	\Lambda_y^\chi	&=	\tau \frac{eB}{c} \frac{\nabla T\frac{\epsilon-\mu}{T}  - eE}{\left( 1 + \frac{e}{c} \vec{B} \cdot \vec{\Omega}^\chi \right)^2 + \omega_c^2 \tau^2} \left[ \Omega_z^\chi \omega_c \tau - \left( 1+ \frac{e}{c} \vec{B} \cdot \vec{\Omega}^\chi \right)^{-1} \Omega_y^\chi \omega_c^2 \tau^2 \right],	\\
	\Lambda_z^\chi	&=	\tau \frac{eB}{c} \frac{\nabla T\frac{\epsilon-\mu}{T}  - eE}{\left( 1 + \frac{e}{c} \vec{B} \cdot \vec{\Omega}^\chi \right)^2 + \omega_c^2 \tau^2} \left[ -\Omega_y^\chi \omega_c \tau - \left( 1+ \frac{e}{c} \vec{B} \cdot \vec{\Omega}^\chi \right)^{-1} \Omega_z^\chi \omega_c^2 \tau^2 \right].
\end{align}
Calculating $\kappa_{xx}$ and $\kappa_{xy}$, we find that the presence of the electric field does not change the result previously obtained in Secs.~\ref{sec:TBPAR} and~\ref{sec:TBPREP}, respectively. Physically, this is a result of the terms linear in chirality canceling out when summing over chiralities and the assumption of linear response. Note that, if inter-nodal scattering cannot be neglected, this correction will generally be non-zero, see. Ref.~[\onlinecite{PhysRevB.89.195137}].

\subsection{$\vec{\nabla}T=\nabla T \hat{x},\vec{B}=B \hat {z},\vec{E}=E\hat{z}$}
When the temperature gradient is perpendicular to the electric and magnetic fields (see Fig.~\ref{FiguresE}(b)), the Boltzmann equation reads
\begin{equation}
	\left( 1+ \frac{e}{c} \vec{B} \cdot \vec{\Omega}^\chi \right)^{-1} \left[ \left( v_x - eE \Omega^\chi_x \right)\nabla T \frac{\partial f^\chi}{\partial T} + eE\frac{\partial f^{\chi}}{\partial p_z} + \frac{e}{c} v_y B \frac{\partial f^{\chi}}{\partial p_x} - \frac{e}{c} v_x B \frac{\partial f^{\chi}}{\partial p_y} + \frac{e^2}{c} \left( \vec{E} \cdot \vec{B} \right) \vec{\Omega}^\chi \cdot \vec{\nabla}_p f^{\chi} \right]	=\frac{f^{\chi}-f_{eq}}{\tau}.
\end{equation}
We assume a solution
\begin{equation}
f^\chi	=	f_{eq} - \left( 1+ \frac{e}{c} \vec{B} \cdot \vec{\Omega}^\chi \right)^{-1} \tau\left( eE \frac{\partial f_{eq}}{\partial p_z} + \frac{e^2}{c} EB \vec{\Omega}^{\chi} \cdot \vec{\nabla}_p f_{eq} + v_x \nabla T \frac{\partial f_{eq}}{\partial T} -eE\Omega^\chi_y \nabla T \frac{\partial f_{eq}}{\partial T}\right) + \left( -\frac{\partial f_{eq}}{\partial \epsilon} \right) \vec{v} \cdot \vec{\Lambda}^\chi.
\end{equation}
It is easy to show that $\Lambda^\chi_z=0$ (again, assuming that internodal scattering can be neglected), but the other two components of $\vec{\Lambda}$ are coupled. Following the solution method outlined in Ref.~[\onlinecite{PhysRevB.89.195137}], we find
\begin{align}
	\Lambda_x^\chi	&= - \tau \frac{e^2}{c}EB \frac{\left( 1+ \frac{e}{c} \vec{B} \cdot \vec{\Omega}^\chi \right)^{-1} \omega_c^2 \tau^2 \Omega_x^\chi - \omega_c \tau \Omega^\chi_y }{ \left( 1+ \frac{e}{c} \vec{B} \cdot \vec{\Omega}^\chi \right)^{2}+ \omega_c^2 \tau^2} +  \chi \tau  \frac{\nabla T \frac{\epsilon-\mu}{T} \frac{e\hbar E}{2v_F p^2} \omega_c\tau}{\left( 1+ \frac{e}{c} \vec{B} \cdot \vec{\Omega}^\chi \right)^{2}+ \omega_c^2 \tau^2}	+ \tau \frac{\nabla T \frac{\epsilon - \mu}{T}  \left( 1+ \frac{e}{c} \vec{B} \cdot \vec{\Omega}^\chi \right)^{-1} \omega_c^2 \tau^2}{ \left( 1+ \frac{e}{c} \vec{B} \cdot \vec{\Omega}^\chi \right)^{2}+ \omega_c^2 \tau^2} ,
\end{align}
\begin{align}
	\Lambda_y^\chi	&= - \tau \frac{e^2}{c}EB \frac{\left( 1+ \frac{e}{c} \vec{B} \cdot \vec{\Omega}^\chi \right)^{-1} \omega_c^2 \tau^2 \Omega_y^\chi + \omega_c \tau \Omega^\chi_x }{ \left( 1+ \frac{e}{c} \vec{B} \cdot \vec{\Omega}^\chi \right)^{2}+ \omega_c^2 \tau^2} +  \chi \tau  \frac{\nabla T \frac{\epsilon-\mu}{T} \frac{e\hbar E}{2v_F p^2} \left( 1+ \frac{e}{c} \vec{B} \cdot \vec{\Omega}^\chi \right)^{-1} \omega_c^2\tau^2 }{\left( 1+ \frac{e}{c} \vec{B} \cdot \vec{\Omega}^\chi \right)^{2}+ \omega_c^2 \tau^2}	+ \tau \frac{\nabla T \frac{\epsilon - \mu}{T}   \omega_c \tau}{ \left( 1+ \frac{e}{c} \vec{B} \cdot \vec{\Omega}^\chi \right)^{2}+ \omega_c^2 \tau^2} .
\end{align}
Substituting this correction back into the ansatz for $f$, and calculating $\kappa_{xx}$ and/or $\kappa_{xy}$ and summing over both chiralities, we find that most terms either cancel from the summation over two Weyl nodes of opposite chirality or vanish due to the angular part of the momentum integration. Keeping only lowest order in $B$, Eq.~\eqref{LtwotwonoEfield} is recovered, implying that the chiral anomaly does not have a significant impact on the thermal transport for parallel electric and magnetic fields that are perpendicular to the thermal gradient.

\begin{table*}
\begin{tabular}{|c|c|c|c|}
\hline
Coefficient &~ $\vec{\nabla}T=\nabla T \hat{x},\vec{B}=B \hat {x},\vec{E}=0\;\&\; \vec{E}=E\hat x$ ~&~ $\vec{\nabla}T=\nabla T \hat{x},\vec{B}=B \hat {z},\vec{E}=0\; \&\; \vec{E}=E \hat z$ & ~ $\vec{\nabla}T=\nabla T \hat{x},\vec{B}=B \hat {y},\vec{E}=E\hat z$ ~\tabularnewline
\hline 
 $\kappa_{xx}$ &  $2\kappa_{xx}(B=0)\left(1+\left(\frac{eB}{c}\right)^2\frac{v_f^4}{\mu^4}\right).$ & $2 \kappa_{xx}(B=0)\left(1-\left(\omega_c^2\tau^2\right)\right) $ &  $2\kappa_{xx}(B=0)\left(1-\left(\omega_c^2\tau^2\right)\right)$~ \tabularnewline
\hline
 $\kappa_{xy}$ & $0$ & $\frac{2\pi}{3h}k_B^2T g B-2 \kappa_{xx}(B=0)\omega_c\tau$ & 0~  \tabularnewline
\hline
\end{tabular}
\caption{The thermal conductivity for various magnetic and electric field configurations. All variables and regimes of validity are given in the main text.}
\label{Fields}
\end{table*}

\subsection{$\vec{\nabla}T=\nabla T \hat{x},\vec{B}=B \hat {y},\vec{E}=E\hat {z}$}\label{sec:ALLPREP}
We now consider the case where all fields are perpendicular (see Fig.~\ref{FiguresE}(c)) for completeness and to highlight the uniqueness of the results in Sec.~\ref{sec:TBPAR}. In this case, the Boltzmann equation reads
\begin{align}
\left(1+\frac{e}{c}\vec{B}\cdot\vec{\Omega}^{\chi}\right)^{-1}\left\{ \left( v_x - eE \Omega^\chi_y \right)\nabla T \frac{\partial f^\chi}{\partial T} +
\left(eE\frac{\partial f^{\chi}}{\partial p_z}+\frac{e}{c} v_x B \frac{\partial f^{\chi}}{\partial p_z} - \frac{e}{c} v_z B \frac{\partial f^{\chi}}{\partial p_x} \right)\right\}=-\frac{f^\chi-f_{eq}}{\tau}.
\label{FullB}
\end{align}
We assume a solution of the form
\begin{equation}
f^\chi	=	f_{eq} - \left( 1+ \frac{e}{c} \vec{B} \cdot \vec{\Omega}^\chi \right)^{-1} \tau\left( -eE v_z+ v_x \nabla T \frac{\epsilon-\mu}{T} -eE\Omega^\chi_y \nabla T\frac{\epsilon-\mu}{T} \right) \left( -\frac{\partial f_{eq}}{\partial \epsilon}\right) + \left( -\frac{\partial f_{eq}}{\partial \epsilon} \right) \vec{v} \cdot \vec{\Lambda}^\chi.
\end{equation}
Solving for $\Lambda$ yields
\begin{equation}
\Lambda^\chi_x=-\tau\left(\frac{\omega_c\tau eE -\omega_c^2\tau^2\nabla T \frac{\epsilon-\mu}{T}\left( 1+ \frac{e}{c} \vec{B} \cdot \vec{\Omega}^\chi \right)^{-1}}{\left( 1+ \frac{e}{c} \vec{B} \cdot \vec{\Omega}^\chi \right)^2+\omega_c^2\tau^2}\right),
\end{equation}
\begin{equation}
\Lambda^{\chi}_z=-\tau\left(\frac{\omega_c\tau\nabla T \frac{\epsilon-\mu}{T}+\omega_c^2\tau^2eE \left( 1+ \frac{e}{c} \vec{B} \cdot \vec{\Omega}^\chi\right)^{-1}}{\left( 1+ \frac{e}{c} \vec{B} \cdot \vec{\Omega}^\chi\right)^2+\omega_c^2\tau^2}\right),
\end{equation}
and $\Lambda^{\chi}_y=0$. From there, we can calculate $L^{12}_{xx}$ and $L^{22}_{xx}$ for the two Weyl nodes. We find
\begin{equation}
L^{12}_{xx}=-2e\int\frac{\mathrm{d}^3p}{(2\pi)^3}v_x^2\tau\frac{\epsilon-\mu}{T} \left(-\frac{\partial f_{eq}}{\partial \epsilon}\right) \left(1-\omega_c^2\tau^2\right),
\end{equation}
\begin{equation}
L^{22}_{xx}=2\int\frac{\mathrm{d}^3p}{(2\pi)^3}v_x^2\tau\frac{(\epsilon-\mu)^2}{T}\left(-\frac{\partial f_{eq}}{\partial \epsilon}\right) \left(1-\omega_c^2\tau^2\right).
\end{equation}
Plugging this into Eq.~\ref{eq:kappa_tensor}, we find the longitudinal thermal conductivity for the two Weyl nodes to be
\begin{equation}
\kappa_{xx}(B)=2\kappa_{xx}(B=0)\left(1-(\omega_c\tau)^2\right).
\end{equation}
Turning to $L^{12}_{xy}$ and $L^{22}_{xy}$, we find, from Eqs. \eqref{Nernst_Weyl} and \eqref{Thermal_Hall_1},
\begin{equation}
L^{12}_{xy}=-\frac{e}{\hbar}\frac{1}{T}\int\frac{\mathrm{d}^3p}{(2\pi)^3}\Omega_z^{\chi}(\epsilon-\mu)f_{eq}=0,
\end{equation}
and 
\begin{equation}
L^{22}_{xy}=-\frac{e}{\hbar}\frac{1}{T}\int\frac{\mathrm{d}^3p}{(2\pi)^3}\Omega_z^{\chi}(\epsilon-\mu)^2f_{eq}=0.
\end{equation}

\end{widetext}

A summary of the thermal conductivity results for all the different field configurations are given in Table~\ref{Fields}.
\section{Comparisons to the Kubo Formula}\label{sec:Kubo}
We start with the standard expression for $L^{22}_{xx}$ in imaginary time,\cite{Mahan} and closely follow Ref.~[\onlinecite{PhysRevLett.108.046602}].  For this section we use natural units ($\hbar=k_B=1$).   In terms of current correlation functions, we have for a single Weyl node (note that in this section $\tau$ is the imaginary time that appears in the Matsubara formalism)
\begin{equation}
L^{22}_{xx}(\omega_n)=\frac{\beta}{\omega_n}\int_{0}^{\beta}\mathrm{d}\tau e^{i\omega_n\tau}\langle T_{\tau} J_{q,x}(\tau)J_{q,x}(0)\rangle.
\label{eq:Kubo}
\end{equation}
We are interested in the case when the Fermi energy is at the Weyl node, thus $J_{q,x}=J_{E,x}$. The current, when $\mu=0$, is given by
\begin{equation}
J_{E,x}(q)=v_f^2\sum_{k} k_x\delta_{\alpha,\gamma} c_{k+\frac{q}{2},\alpha}^{\dagger}c_{k-\frac{q}{2},\gamma}.
\label{eq:current}
\end{equation}
The current is in the same form as the free case because interactions simply renormalize the Fermi velocity. \cite{PhysRevB.88.045108} Plugging Eq.~\eqref{eq:current} into Eq.~\eqref{eq:Kubo} with $q\rightarrow0$, applying Wicks theorem, and keeping only connected diagrams, we arrive at
\begin{equation}
L^{22}_{xx}=\beta\frac{v_f^4}{\omega_n}\int_{0}^{\beta}\mathrm{d}\tau e^{i\omega_n\tau}\sum_{k}k_x^2 \delta_{\alpha,\gamma}\delta_{\rho,\theta} G(\tau,k)_{\alpha,\theta}G(-\tau,k)_{\gamma,\rho},
\end{equation}
Switching to the helical basis and performing the trace, as done in Ref.~[\onlinecite{PhysRevLett.108.046602}], we have
\begin{equation}
L^{22}_{xx}=2\beta\frac{v_f^4}{\omega_n}\int_{0}^{\beta}\mathrm{d}\tau e^{i\omega_n\tau}\sum_{\lambda,\lambda'}\sum_{k}k_x^2 G_{\lambda}(\tau,k)G_{\lambda'}(-\tau,k)
\label{L22}
\end{equation}
Introducing the Fourier transform of the Green's function
\begin{equation}
G(\tau)=\sum_{m}e^{i\omega_m\tau}G(i\omega_m), 
\label{FTG}
\end{equation}
and inserting Eq.~\eqref{FTG} into Eq.~\eqref{L22} and performing the $\tau$ integral, we obtain
\begin{equation}
L^{22}_{xx}=2\beta\frac{v_f^4}{\omega_n}\sum_m\sum_{\lambda,\lambda'}\sum_{k}k_x^2 G_{\lambda}(i\omega_n+i\omega_m,k)G_{\lambda'}(i\omega_m,k) .
\end{equation}
After performing the standard Matsubara sum, using rotational symmetry, and analytically continuing to real frequencies, we take the limit as $\omega\rightarrow0$ and find
\begin{equation}
L^{22}_{xx}=\frac{v_f^4}{3}\frac{\beta}{(\pi)^3}\int\mathrm{d}\epsilon(-\frac{\partial f}{\partial \epsilon}) I(\epsilon),
\label{L_22_Pre}
\end{equation}
where the Green's function is
\begin{equation}
G_{\lambda}(\epsilon,k)=\frac{1}{\epsilon-\lambda v_f k-\Sigma},
\end{equation}
$\Sigma$ is the self-energy, and 
\begin{equation}
I(\epsilon)=\sum_{\lambda,\lambda'}\int_0^{\Lambda/v_f}\mathrm{d}k k^4 \mathrm{Im}G_{\lambda}(\epsilon,k) \mathrm{Im}G_{\lambda'}(\epsilon,k).
\end{equation}
Here $\Lambda \propto T$ is a large momentum cut-off that needs to be introduced since the integral is divergent as $\Lambda\rightarrow\infty$. The cutoff is proportional the the temperature since it is the only energy scale in the problem. Taking the imaginary part of the Green's function, we find
\begin{equation}
\mathrm{Im}G_{\lambda}(\epsilon,k)=\frac{\mathrm{Im}\Sigma}{(\epsilon-\lambda v_fk-\mathrm{Re}\Sigma)^2+\mathrm{Im}\Sigma^2}.
\end{equation}
As stated before, the real part of the self energy simply renormalizes the Fermi velocity and $\mathrm{Im}\Sigma=\frac{\alpha^2T}{2}$ (See Section \ref{sec:TTC} C. We have set $A=1$ for simplicity). In the limit of large $T$, we have
\begin{equation}
I(\epsilon)=\frac{\pi}{2}\frac{\epsilon^4}{\alpha^2T}.
\end{equation} 
Plugging this into Eq.~\eqref{L_22_Pre}, we obtain
\begin{equation}
L^{22}_{xx}=\frac{7\pi^2}{90}\frac{T^2}{\alpha^2v_f}.
\end{equation}
This is in exact agreement with the Boltzmann equation result when $A,k_B,\hbar$ are set to one.

We close this section by noting that when the Fermi energy is at the Weyl node, the Kubo formula gives $L^{12}=0$, in agreement with the Boltzmann equation. Physically, this is due to particle-hole symmetry.

\section{Conclusions}\label{sec:Con}

In our work, we have analytically investigated the electronic contribution to the thermoelectric properties of Dirac and Weyl semimetals via the Boltzmann equation. We considered the cases where transport is relaxed by disorder and electron-electron interactions. We find an interesting dependence of the  thermoelectric coefficients on the Fermi energy {(\it i.e.} doping away from the Dirac or Weyl point).  Notably, in the case of interactions we find that the longitudinal thermal conductivity has an interesting quadratic temperature dependence, in contrast to a linear dependence on the temperature for scattering from charged impurities that dope the system or short-range electrically neutral disorder.  A linear temperature dependence of the thermal conductivity is the expected result for a ``generic" metallic system (one without Weyl or Dirac points).  We stress that in our work we have ignored the contribution from phonons to the thermal conductivity, which are expected to dominate at high enough temperatures.  The lattice contributions are less generic than the electronic one, so we leave their study to other work.

We have also considered the effect of electric and magnetic fields on the thermoelectric coefficients. Notably, when the magnetic field and temperature gradient are parallel we find a large positive contribution to the longitudinal thermal conductivity that is quadratic in magnetic field strength, similar to the magnetic field dependence of the longitudinal electrical conductivity due to the presence of the chiral anomaly when there is no thermal gradient present, and there is a vanishing transverse thermal Hall conductivity. When the magnetic field is perpendicular to the temperature gradient, we find that the thermal conductivity is linear in magnetic field strength, and the longitudinal thermal conductivity picks up a negative contribution that goes as the square of the magnetic field. The presence of electric fields does not change these results under the assumption of no inter-node scattering. We also calculated the thermal conductivity via the Kubo formula for the case of interactions and find exact agreement with our Boltzmann equation results at high temperatures. Taken together, our theoretical results provide some concrete experimental tests that can be usefully applied in the search for three-dimensional systems with a linear electron dispersion.  Our main results are summarized in Table~\ref{Nofields} and Table~\ref{Fields} of the main text.

We note that it would be interesting to study the effect of system size and the contribution from Fermi arcs, as recently edge states and tuning the system sizes have been shown to improve thermoelectric performance in topological insulators. \cite{PhysRevLett.112.226801} It would also be worthwhile to investigate the effects of relaxation via a slow imbalance of carriers as was done in graphene.\cite{PhysRevB.79.085415} We hope our paper motivates the experimental study of the thermoelectric properties of Dirac and Weyl semimetals, as well as further theoretical work on the subject.

{\bf \em Acknowledgments} -- RL thanks V. Chua, V. Zyuzin, B. Sbierski, D. Lorshbough, and C. Br\"{u}ne for useful discussions. We thank W. Witczak-Krempa for valuable comments and suggestions. RL was supported by NSF Graduate Research Fellowship award number 2012115499. RL is grateful for the hospitality of the Institute for Theoretical Physics, University of W\"{u}rzburg under European Research Council grant ERC- StG-Thomale-2013-336012, during the beginning stages of this work. PL and GAF acknowledges financial support through ARO Grant No. W911NF-09-1-0527 and NSF Grant No. DMR-0955778.

\appendix
\section{Self-energy for a Disordered Weyl Metal in the Self-consistent Born Approximation}
Ref.~[\onlinecite{PhysRevB.89.014205}] calculated the self-energy for the case of a slowly varying background disorder potential using the self-consistent Born approximation, finding
\begin{equation}
	\Sigma \left( \epsilon, \vec{k} \right) = C \left( \epsilon -\Sigma \right) \left[ -v\Lambda \mp i \left( \epsilon -\Sigma \right) \frac{\pi}{2} \right] + O \left( \frac{\epsilon}{\Lambda} \right),
\end{equation}
where $C=\frac{\gamma}{2\pi^2 (\hbar v_f)^3}$, and $\Lambda$ is a momentum cutoff. We now solve the quadratic equation for the self energy for $\epsilon \ll \Lambda$ and $Cv_f\Lambda < 1$. Series expanding the imaginary part of the self energy in terms of quasiparticle energy yields (to lowest order in disorder strength, $\gamma$)
\begin{equation}
	\mathrm{Im} \Sigma=\frac{\pi}{2}C \epsilon^2  \left[ 1-\frac{5\pi^2}{4} C^2  \epsilon^2\right] + O \left( \epsilon^6 \right),
\end{equation}
where we have picked the root with no term in the imaginary part constant in $\epsilon$, since that would correspond to a finite lifetime when all energy scales are set to zero. The second order correction to the imaginary part of the self energy was not considered by Ref.~[\onlinecite{PhysRevB.89.014205}] and BHB, and is crucial in obtaining a non-zero thermopower when the Fermi energy lies away from the Weyl node. The scattering time is then
\begin{equation}
\tau=\frac{3}{2}\frac{1}{2\mathrm{Im} \Sigma}\approx\frac{3}{2}\frac{1}{2\pi\gamma g(\epsilon)} \left(1+\frac{5}{16\pi^2}\frac{\gamma^2}{\hbar^6v_f^6}\epsilon^2\right).
\end{equation}

\section{An Expression for Energy Current}
In this section, we derive an expression for the energy current for free particles. We begin with the continuity equation (in real space)
\begin{equation}
\frac{\partial h(\vec{x})}{\partial t}+\vec{\nabla}\cdot \vec{J}_E(\vec{x})=0.
\end{equation}
For simplicity, we take $h(\vec{x})$ to be a simple hopping model with translational invariance, given by
\begin{equation}
h(\vec{x})=\sum_{\vec{y}}h_{\vec{x}-\vec{y}}\left(c^{\dagger}_{\vec{x}} c^{
\phantom{\dagger}}_{\vec{y}}+c^{\dagger}_{\vec{y}} c^{\phantom{\dagger}}_{\vec{x}}\right).
\end{equation}
The full Hamiltonian of the system is 
\begin{equation}
H=\sum_{\vec{x}} h(\vec{x}).
\end{equation}
Taking the Fourier transform of $h(\vec{x})$, we have
\begin{equation}
h(\vec{x})=\sum_{\vec{k},\vec{q}}h_k\left( c^{\dagger}_{\vec{k}-\vec{q}}c^{\phantom{\dagger}}_{\vec{k}}+c_{\vec{k}}^{\dagger}c^{\phantom{\dagger}}_{\vec{k}+\vec{q}}\right) e^{i\vec{q}\cdot\vec{x}}.
\end{equation}
The current conservation equation (in momentum space) then becomes
\begin{equation}
\frac{\partial h(\vec{q})}{\partial t}-i \vec{q}\cdot\vec{J}_E=0.
\end{equation}
We then use the Heisenberg equation of motion,
\begin{equation}
\frac{\partial h(\vec{q})}{\partial t}=-i[h(\vec{q}),H],
\end{equation}
to solve for $J_E$. After some algebra and allowing for spin indices, we find
\begin{equation}
\vec{J}_{E}(\vec{q})=\frac{1}{2}\sum_{k}\left(\frac{\partial H^{\alpha\beta}}{\partial \vec{k}}H^{\beta\gamma}+ H^{\alpha\beta}\frac{\partial H^{\beta\gamma}}{\partial \vec{k}}\right)c_{\vec{k}+\vec{\frac{q}{2}},\alpha}^{\dagger}c^{\phantom{\dagger}}_{\vec{k}-\vec{\frac{q}{2}},\gamma}.
\end{equation}
Using the Hamiltonian in the main text ($H_{\alpha,\beta}=\hbar v_f \vec{\sigma}_{\alpha,\beta}\cdot \vec{k}-\mu\delta_{\alpha,\beta}$), we have
\begin{align}
\frac{1}{2}\left(\frac{\partial H^{\alpha\beta}}{\partial k_x}H^{\beta\gamma}+ H^{\alpha\beta}\frac{\partial H^{\beta\gamma}}{\partial k_x}\right)=\nonumber \\
\hbar^2v_f^2k_x\delta_{\alpha,\gamma}-\frac{\hbar v_f \mu}{2} \left(\sigma_{\alpha,\beta}^x\delta_{\beta,\gamma}-\delta_{\alpha,\beta}\sigma_{\beta,\gamma}^x\right)
\end{align}
so finally
\begin{align}
J_{E,x}(\vec{q})=\sum_{k} (\hbar^2v_f^2 k_x\delta_{\alpha,\gamma}- \nonumber \\
\frac{\hbar v_f \mu}{2} \left.\left(\sigma_{\alpha,\beta}^x\delta_{\beta,\gamma}-\delta_{\alpha,\beta}\sigma_{\beta,\gamma}^x\right)\right)c_{\vec{k}+\vec{\frac{q}{2}},\alpha}^{\dagger}c^{\phantom{\dagger}}_{\vec{k}-\vec{\frac{q}{2}},\gamma}.
\end{align}
This is the expression we need to calculate the current-current correlation function which determines the thermal conductivity.

\bibliography{outline}

\end{document}